\input a4.sty
\equano 0
\def\chapter#1{\goodbreak\vskip 10mm\vglue 5mm
\centerline{\fontetitre #1}\nobreak\vskip 15mm\global\sectno 0}
\font\rmtw cmr10 at 12pt

\vglue 15mm
\centerline{\bftw QCD ET LES PROCESSUS EXCLUSIFS DURS\footnote{*} {\sl cours
 donn\'e \`a la $27^{\`eme}$ Ecole
d'Et\'e de Physique des Particules, Clermont-Ferrand, France (Septembre
1995).}}
\vskip 1.5cm
\centerline{\rmtw Thierry GOUSSET$^{1,2}$ et Bernard PIRE$^{2}$}
\vskip 1cm
\centerline{\it 1. CEA, Service de Physique Nucl\'eaire/DAPNIA, CE Saclay,
 F91191 Gif, France}

\centerline{\it 2. Centre de Physique Th\'eorique, Ecole Polytechnique,
 F91128 Palaiseau, France}

\vskip 1cm

{\sl Avertissement: on utilise les notations d'Itzykson et Zuber~{\rm
[Itzykson]},
sauf pour la
normalisation des \'etats fermioniques que l'on choisit identique \`a celle
des \'etats bosoniques. Il s'ensuit que les spineurs ont la normalisation
$\bar{u}u=2m$. Les sommations sur les indices r\'ep\'et\'es sont implicites.}

\medskip
Nous consid\'erons ici les {\it processus
 exclusifs}, c'est \`a dire
les interactions \`a deux corps dans l'\'etat initial (ou la
d\'esint\'egration d'une particule instable) r\'esultant en un \'etat final
o\`u tous les produits sont identifi\'es. L'utilisation du d\'eveloppement
perturbatif n'est {\it a priori} envisageable que pour une r\'eaction dite
{\it dure}, c'est \`a dire impliquant un grand transfert d'impulsion.

Pr\'ecisons ceci sur un exemple: la diffusion Compton sur le nucl\'eon. Le
processus est
$$
\gamma(k)+N(p)\rightarrow\gamma(k')+N(p'),
$$
et l'on
s'int\'eresse \`a la section efficace diff\'erentielle, polaris\'ee ou non,
$$
{d\sigma\over dt}(s,t),\ s=(k+p)^2,\ t=(k-k')^2,
$$
pour les grandes valeurs de $|t|\sim s$ (au regard desquelles les \'echelles
hadroniques du confinement peuvent \^etre n\'eglig\'ees).
A la limite ultra-relativiste, en n\'egligeant la masse du nucl\'eon,
la cin\'ematique se simplifie. Dans le r\'ef\'erentiel du centre de masse,
on a les relations
$$
E={\sqrt{s}\over 2},\ \sin^2{\theta\over 2}=-{t\over s},
$$
o\`u $E$ est l'\'energie de chacune des particules et $\theta$ l'angle entre
le photon incident et celui \'emergeant. Ainsi parle-t-on aussi
d'interaction \`a grand angle et grande \'energie.

Les premi\`eres propri\'et\'es qui ont \'et\'e \'etudi\'ees pour ces
processus sont les {\it r\`egles de comptage} qui permettent de pr\'edire,
\`a grand angle fix\'e, la loi de puissance dans l'\'energie
$\sqrt{s}$ selon laquelle leur section efficace d\'ecro\^it~[Bro73].
Ces r\`egles
sont contemporaines \`a l'\'elaboration de QCD et ne sont par ailleurs pas
propres \`a cette th\'eorie.

\chapter{A- Les r\`egles de comptage}

Il existe deux moyens standards de les envisager: un raisonnement qui
s'appuie sur le d\'eroulement spatiotemporel des r\'eactions et un
argument dimensionnel. Ils reposent tous deux sur l'hypoth\`ese que
le m\'ecanisme \'el\'ementaire est dur, c'est \`a dire que {\it tous} les
constituants \'el\'ementaires subissent un grand transfert d'impulsion au
cours du processus. La justification rigoureuse de la validit\'e de cette
hypoth\`ese \`a partir de la th\'eorie des champs n'est pas ais\'ee.
On reviendra sur cette difficult\'e dans le deuxi\`eme chapitre.
On peut, en attendant, adopter un point de vue pragmatique et parler de
pr\'edictions \'etablies au moyen de QCD perturbative dans le mod\`ele des
sous processus durs.

\section{ Image spatiotemporelle: Facteur de forme:}

La quantit\'e la plus simple parmi les observables exclusives
est le {\it facteur de forme du pion}.

On consid\`ere le processus $e^-\pi^+\rightarrow e^-\pi^+$. L'interaction
\'electromagn\'etique est v\'ehicul\'ee par l'\'echange de photons
virtuels; les effets dus \`a l'\'echange
de plusieurs photons, d'ordre $\alpha_{em}$ relativement \`a l'\'echange
d'un seul photon, sont n\'egligeables.
On se limite donc \`a l'\'echange d'un seul photon comme represent\'e sur
la Figure~1.

\midinsert
\vskip 3cm
\centerline{\sl Figure 1: Facteur de forme du pion}
\endinsert

Le pion est une particule pseudoscalaire. Si elle \'etait \'el\'ementaire,
la section efficace de diffusion serait
$$
\left.{d\sigma\over dt}\right|_{\rm point}={4\pi\alpha^2\over t^2}
{(s-m^2-M^2)^2+t(s-m^2)\over\left(s-(m+M)^2\right)\left(s-(m-M)^2\right)},
$$
o\`u $s=(k+p)^2$ et $t=(k-k')^2=-Q^2\le 0$. $m$ et $M$ sont les masses de
l'\'electron et du pion.

En fait le pion est un objet composite et la section efficace s'\'ecrit
$$
{d\sigma\over dt}=|F_{\pi}(Q^2)|^2\,\left.{d\sigma\over dt}\right|_{\rm point},
\num
$$
ce qui d\'efinit le facteur de forme du pion $F_\pi$. Il mesure la capacit\'e
du pion \`a rester intact lors de l'impact avec l'\'electron. C'est  une
quantit\'e tr\`es sensible aux m\'ecanismes de confinement puisqu'on
assiste de fait
\`a la restauration de l'int\'egrit\'e du m\'eson apr\`es que le choc avec
l'\'electron
ait bouscul\'e l'ordonnancement des quarks. A la limite
$Q^2=0$, lorsque la structure du pion n'est pas r\'esolue, on a $F_{\pi}(0)=1$.

La param\'etrisation de l'Eq.(\the\equano) se d\'emontre en \'ecrivant
l'\'el\'ement de matrice~$S$ sous la forme
$$\langle e\pi|S|e\pi\rangle=\int d^4xd^4y
\langle\pi|J^{\mu}(x)|\pi\rangle\langle0|T(A_{\mu}(x)A_{\nu}(y))|0\rangle
\langle e|j^{\nu}(y)|e\rangle,
$$
o\`u $J^{\mu}$ et $j^{\nu}$ sont respectivement les courants
\'electromagn\'etiques des quarks et de l'\'electron. On a ainsi isol\'e
l'\'el\'ement de matrice o\`u le pion intervient
$$
\langle\pi^+(p')|J^{\mu}(x)|\pi^+(p)\rangle=
\langle\pi^+(p')|J^{\mu}(0)|\pi^+(p)\rangle e^{i(p'-p)x}
$$

Comme le pion est une particule (pseudo-) scalaire la param\'etrisation la plus
g\'en\'erale de l'\'el\'ement de matrice ci-dessus (un quadrivecteur) doit
\^etre envisag\'ee \`a l'aide des quadrivecteurs $(p+p')^{\mu}$ et
$(p'-p)^{\mu}$ pond\'er\'es par des fonctions de $Q^2$, le seul scalaire du
probl\`eme (la masse du pion, $m_{\pi}$, mise \`a part). Par ailleurs, le
courant \'electromagn\'etique est conserv\'e: $\partial_{\mu}J^{\mu}=0$; il
s'ensuit qu'un terme en $(p'-p)^{\mu}$ est n\'ecessairement nul, d'o\`u
$$
\langle\pi^+(p')|J^{\mu}(0)|\pi^+(p)\rangle=e_{\pi}(p+p')^{\mu}F_{\pi}(Q^2).
\num
$$
\newcount\parametrisation\parametrisation\the\equano
Notons enfin que l'hermiticit\'e du courant entraine la r\'ealit\'e du
facteur de forme (pour une transition de genre espace, conf\`ere B-5) et que,
pour une particule ponctuelle ($F(Q^2)=1$), on
retrouve le terme de couplage de l'\'electrodynamique scalaire.

\smallskip
C'est sur ce facteur de forme $F_{\pi}(Q^2)$ que l'on peut le plus
facilement deviner la d\'ependance en $Q^2$ (au grande valeur de cette
variable) en examinant la mani\`ere dont se d\'eroule le processus.
A cette fin, on se place dans le centre de masse de la r\'eaction, et
l'on consid\`ere le cas, illustr\'e sur la Figure~2, o\`u l'\'electron final
\'emerge avec un angle de 180$^{\circ}$ par rapport \`a l'\'electron initial
(les autres angles pourvu qu'ils ne soient pas trop petits conviennent aussi
pour l'argument qui suit).

\midinsert
\vskip 3cm
\centerline{\sl Figure 2:Image spatiotemporelle du processus
$e^-\pi^+\rightarrow e^-\pi^+$}
\endinsert

Dans son r\'ef\'erentiel de repos le pion est repr\'esent\'e comme une
collection de partons, les quarks et les gluons, {\it grosso-modo}
uniform\'ement r\'epartis dans une sph\`ere de rayon $R_{\pi}$ (typiquement
le rayon de charge du pion, soit environ $0.5\,$fm). Dans le centre de
masse de la r\'eaction, la dimension
longitudinale est contract\'ee en une \'epaisseur $R_{\pi}/\gamma$ avec
$\gamma=Q/2M$. Les dimensions transverses, elles, ne sont pas
affect\'ees par la contraction de Lorentz. A l'instant 0, l'\'electron frappe
un des partons, dit parton {\it actif}, et tous deux rebroussent chemin. Pour
que le processus complet soit \'elastique, {\it tous} les autres partons
doivent \^etre pr\'evenus de l'accident avant l'instant $t\approx1/Q$ pour
former le pion \'emergeant (aussi contract\'e dans notre r\'ef\'erentiel
d'\'etude). Le mouvement du parton actif est apr\`es la collision
$z(t)=-t,\ x(t),y(t)=0$ alors que celui d'un {\it spectateur} est
$z(t)=t+z_0,\ x(t)=x_0,\ y(t)=y_0$ (on a $-1/Q\lapprox z_0\lapprox1/Q$ et
$-R\le x_0,y_0\le R$). Entre les instants 0 et $1/Q$ un parton spectateur ne
peut recevoir un signal physique \'emis par le parton actif \`a l'instant 0
que si l'intervalle $\Delta=t^2-(t+z_0)^2-x_0^2-y_0^2$ est positif, soit
si il est situ\'e \`a une distance $\sqrt{x_0^2+y_0^2}\le1/Q$ dans le plan
transverse. On est donc amen\'e \`a compter la probabilit\'e de trouver
les partons spectateurs, tant dans l'\'etat initial que dans l'\'etat final,
dans un disque transverse de rayon $1/Q$. On trouve
$$
F_{\pi}^2\propto\left({\pi Q^{-2}\over \pi R^2_{\pi}}\right)
^{n_{in}-1+n_{out}-1}.
\num
$$

Dans un pion se trouvent au moins le quark et l'antiquark de valence, cet
\'etat particulier, tant initial que final, contribue en $1/Q^2$. Ajoutons,
par exemple un gluon \`a la valence dans l'\'etat initial, sans changer
l'\'etat final, cette combinaison contribuera pour $1/Q^3$ \etc\ Ces
contributions diminuent au fur et \`a mesure que l'\'energie augmente
relativement \`a celle de l'\'etat de valence.

Une caract\'eristique essentielle se d\'egage imm\'ediatement de cette
\'etude des facteurs de forme et sera par la suite g\'en\'eralis\'ee aux
autres r\'eactions exclusives: {\sl lorsque l'interaction est \`a courte
distance, c'est l'\'etat de valence qui apporte la contribution dominante
en terme de loi d'\'echelle}. De plus, et ce sera la base du concept de {\it
transparence de couleur}, les configurations du m\'eson qui contribuent sont
celles de petites tailles transverses ($O(1/Q)$).

R\'esumons: asymptotiquement, on pr\'edit pour les facteurs de forme,
respectivement du pion et du nucl\'eon, les lois de puissance
$$
F_{\pi}(Q^2)\propto {1\over Q^2},\hskip 2cm
F_N(Q^2)\propto {1\over Q^4}.
\num
$$
Dans le cas du proton, il existe deux facteurs de forme et le raisonnement
d\'evelopp\'e ici n'est pas suffisant pour les distinguer. En fait, si on
s\'epare les facteurs de forme suivant le degr\'e de conservation de
l'h\'elicit\'e,
on montre que la r\`egle de comptage s'applique pour les processus conservant
l'h\'elicit\'e (et donc pour le facteur de forme magn\'etique $G_M$) mais
qu'une
suppression suppl\'ementaire affecte $G_E$.

\section{ Argument dimensionnel: Diffusion Compton}

L'argument qui suit est g\'en\'eral, mais on va l'expliciter sur le processus
d\'ej\`a discut\'e
$$
\gamma+N\rightarrow\gamma+N.
$$

De mani\`ere g\'en\'erale, la diffusion $AB\rightarrow CD$ a, \`a la
limite ultra relativiste, la section efficace diff\'erentielle:
$$
{d\sigma\over dt}={1\over 16\pi s^2}|{\cal M}|^2,\num
$$
\newcount\sectionefficace\sectionefficace\the\equano
que l'on consid\`ere dans la suite \`a $t/s=O(1)$ fix\'e. Il nous faut, pour
trouver la loi d'\'echelle de la r\'eaction, identifier la puissance de $s$
de l'amplitude ${\cal M}$.

L'amplitude ${\cal M}$ est calcul\'ee au moyen des r\`egles de Feynman et on
peut {\it a priori} identifier les dimensions, en unit\'e d'\'energie, des
diff\'erentes quantit\'es qui constituent ces r\`egles:
$$
\vbox{
\halign{#\tvi\hrulefill&\hfil#\cr
-- un spineur externe a une dimension &$1/2$,\cr
-- un vecteur externe &0,\cr
-- un propagateur de fermion &$-1$,\cr
-- un propagateur de boson &$-2$,\cr
-- le vertex boson-fermions &0,\cr
-- le vertex \`a 3 gluons &1,\cr
-- celui \`a 4 gluons &0.\cr
}}$$

Construisons alors un graphe {\it en arbre et connexe} du processus au
niveau des particules \'el\'ementaires (voir Figure~3) et comptons la dimension
obtenue avec les correspondances ci-dessus: on trouve $-4$. On se convainc
ais\'ement que la dimension obtenue par cette d\'emarche ne d\'epend pas de
l'\'eventuelle insertion de boucles et que, pour
tout graphe connexe, la dimension ne d\'epend que du nombre de pattes externes,
 $N$, et vaut $4-N$.

\midinsert
\vskip 3cm
\centerline{\sl Figure 3: Un graphe connexe contribuant \`a la diffusion
Compton}
\endinsert

Dans le calcul de ${\cal M}$, il faut trouver les impulsions de chacune des
lignes puis expliciter les produits scalaires entre ces diverses impulsions
engendr\'es par les r\`egles de Feynman. On se rend alors compte que si l'on
distribue \`a chaque quark ou gluon une fraction {\it finie} de l'impulsion du
hadron auquel il appartient, toutes les particules subissent un grand transfert
d'impulsion (pourvu que le transfert global soit suffisamment grand). Par
suite, tous les produits scalaires sont d'ordre $s$, car c'est la seule
\'echelle dimensionn\'ee pertinente qui r\'egisse cette cin\'ematique.

\smallskip
Dans nos conventions, la dimension compl\`ete de ${\cal M}$ est nulle. Il
manque, en effet, une r\`egle dans la liste ci-dessus pour indiquer comment le
hadron exhibe son contenu en quarks et gluons et la transition du hadron
\`a un
syst\`eme de $n$ partons
$$
|\hbox{Hadron}\rangle\leftrightarrow f_{H,n}|n\hbox{ partons}\rangle
$$
introduit une constante $f_{H,n}$ de dimension $n-1$. Cette transition,
ind\'ependante du processus dur envisag\'e, est r\'egie par le confinement. Il
s'ensuit que l'\'echelle d'\'energie\footnote{$^{\dagger}$}{$M$ d\'esignera
dans la suite, sauf mention explicite du contraire, cette \'echelle des
ph\'enom\`enes de basse \'energie. $M$ peut tout \`a la fois repr\'esenter la
constante de QCD, $\Lambda$, la masse du m\'eson $\rho$ ou l'impulsion
transverse typique des constituants du pion,
$\sqrt{\langle{\bf k}_{\bot}^2\rangle}$, et vaut quelques centaines de MeV.}
naturelle pour mesurer $f_{H,n}$ est ind\'ependante de $s$. Prenant en compte
les transitions hadron-partons, on trouve
$$
{\cal M}=f_{H,n}\sqrt{s}^{\,4-n-1-n'-1}f_{H,n'},
$$
qui est bien sans dimension.

On peut rassembler les consid\'erations pr\'ec\'edentes pour pr\'edire que la
diffusion Compton a un comportement \`a grand angle et grande \'energie
$$
{d\sigma\over dt}\sim {1\over  s^6} f({t\over s}),\num
$$
qui n'est d\^u qu'\`a la {\it transition entre \'etats de valence}.
L'\'etude ci-dessus nous indique, en effet, que le sous processus
$qqqg\gamma\rightarrow qqq\gamma$ contribue \`a la section efficace
diff\'erentielle \`a hauteur de
$$
{f^2_{N,{\rm val}+g}f^2_{N,{\rm val}}\over s^7},
$$
qui est n\'egligeable \`a grande \'energie.

L'amplitude  est donc
s\'epar\'ee, \`a grand transfert, selon
$$
A(s,t/s)=A^{\rm LT}\left(1+O(M/\sqrt{s})\right),\ t/s=O(1)
$$
soit en la contribution $A^{\rm LT}$, dite ``Leading Twist'', c'est \`a dire
de plus petite puissance en $s^{-1}$, et un reste supprim\'e \`a grande
\'energie. On va chercher dans la suite \`a \'evaluer cette contribution
dominante qui, comme on l'a vu, ne n\'ecessite que les quarks de valence
au niveau du sous processus dur.

\section{Le cas exceptionnel du processus de collisions ind\'ependantes}

Cette c\'el\`ebre exception aux r\`egles de comptage a \'et\'e d'abord
remarqu\'ee par P.V. Landshoff~[Lan74]. Les processus de collisions
ind\'ependantes,
appel\'es aussi processus de Landshoff ne respectent pas ce qui vient d'\^etre
\'enonc\'e \`a cause de la pr\'esence de graphes non connexes au niveau des
partons.
Un tel graphe est repr\'esent\'e sur la Figure~4 pour la diffusion $\pi$-$\pi$
:

\midinsert
\vskip 3cm
\centerline{\sl Figure 4: Le  processus de Landshoff}
\endinsert
\noindent
Comme son nom l'indique, le procesus de collisions ind\'ependantes consid\`ere
le cas o\`u, par exemple, les quarks $u$ diffusent \'elastiquement \`a
grand angle de fa\c con ind\'ependante des quarks $\bar d$.
L'analyse dimensionnelle de ce type de contribution se fait comme suit:
les deux quarks qui \'emergent des r\'eactions dures doivent avoir des
directions suffisamment proches pour s'ins\'erer dans des m\'esons finals.
Cette contrainte est exprim\'ee par les fonctions d'onde d\'efinies comme
restreignant les impulsions transverses relatives $k_T$ \`a des valeurs bien
inf\'erieures \`a l'\'echelle $Q$. On peut de fait supposer que  $k_T$ est
pratiquement nul. Comme chaque amplitude de collision se comporte comme
$$
g^2 \bar u u  \bar u u /t
$$
l'\'el\'ement de matrice de la diffusion va comme
$$
[g^2 \bar u u  \bar u u /t ]^{n/4} \sim g^{n/2}
$$
(\`a des corrections logarithmiques pr\`es).
Le comportement de la section efficace vient donc de la contrainte
sur la r\'egion d'int\'egration exprim\'ee par  $\delta^4(
k^1+k^2-k^3-k^4)$. Chaque diffusion met en jeu trois grandes composantes
d'impulsion tandis que la composante orthogonale au plan de diffusion
n'est pas de l'ordre de $\sqrt{s}$ mais plut\^ot d'ordre hadronique, soit
$C<k_T^2>^{1/2}$ , qu'on peut \'ecrire comme $C/<b^2>^{1/2}$, o\`u $b$
est une s\'eparation transverse.

Chaque fonction $\delta$  d'une grande composante compte pour $1/\sqrt{s}$,
puisque
$$
 \delta(p-p') \sim s^{-1/2}\delta(x-x'),
$$
o\`u $x$ et $x'$ sont sans  dimension.
L'amplitude de probabilit\'e  pour qu'une paire de quarks aient des directions
qui coincident suffisamment pour fabriquer un hadron  d\'epend de l'\'energie
comme le  produit des fonctions $\delta$, soit comme $C<b^2>^{1/2} (s)^{-3/2}$.
A l'aide de l'\'equation (\the\sectionefficace), on obtient
$$
{d\sigma\over dt}\propto<b^2> s^{-5},
\num
$$
dans le cas de la diffusion m\'eson-m\'eson.  Quand $s\rightarrow \infty$,
cela l'emporte sur le processus type ``r\`egle de comptage'' qui d\'ecroit
comme
$s^{-6}$.

Dans le cas de la diffusion proton-proton \'elastique, l'argument est identique
mais requiert qu'une autre diffusion quark-quark  coincide en direction avec
les
deux premi\`eres. Cela ajoute trois fonctions $\delta$ de grandes
composantes, si bien
que l'amplitude au carr\'e a un facteur $s^{-3}$ suppl\'ementaire. On a
donc pour
$pp\longrightarrow pp$ :
$$
{d\sigma\over dt}\propto(<b^2>)^2 s^{-8},
\num
$$
qui l'emporte sur le processus ``r\`egle de comptage'' qui d\'ecroit comme
 $s^{-10}$.
\footnote{*} {Le  processus de collisions ind\'ependantes a eu une histoire
confuse, les subtilit\'es n'ayant \'et\'e que graduellement appr\'eci\'ees.
On y reviendra plus bas. Disons seulement pour l'instant qu'il est, de l'avis
des auteurs, \'etabli de fa\c con assez convaincante que ce m\'ecanisme
joue un r\^ole important pour la collision proton-proton dans les  r\'egions
accessibles actuellement, et que l'\'etude du ph\'enom\`ene de transparence
de couleur (voir chapitre E) ne peut se passer de sa contribution. Il faut
 n\'enmoins  reconna\^\i tre que cette opinion est encore l'objet de d\'ebats
dans la communaut\'e.}

\chapter{B- Calcul du facteur de forme du pion}

On veut aller plus loin que la seule donn\'ee de cette loi de puissance et
trouver quantitativement la pr\'ediction de QCD pour le facteur de forme \`a
grand transfert~[Far79]. Ceci nous am\`ene \`a pr\'eciser, d'une part la
fonction d'onde des hadrons et les amplitudes dures de Born, puis le
tra\^itement
des corrections radiatives afin de s'assurer de la validit\'e d'une
factorisation
entre un terme non perturbatif sensible \`a la dynamique du confinement et une
amplitude dure contr\^ol\'ee par un d\'eveloppement perturbatif (am\'elior\'e
par le groupe de renormalisation). Cette factorisation, cruciale pour la
compr\'ehension th\'eorique des \'eventuelles futures donn\'ees
exp\'erimentales,
se dessine sous la forme montr\'ee sur la Figure 5.

\midinsert
\vskip 3cm
\centerline{\sl Figure 5: La factorisation d'un processus exclusif dur:
$X*T_H*X'$}
\endinsert

On se restreint dans ce chapitre au cas p\'edagogique du facteur de forme
du m\'eson $\pi$ mais la technique d\'evelopp\'ee ici est exemplaire de toutes
les r\'eactions exclusives dures.

\section{Comment d\'ecrire le pion?}

On se place dans le r\'ef\'erentiel de Breit o\`u les impulsions sont
$$
q=\pmatrix{0\cr0\cr0\cr Q\cr},\
p=\pmatrix{Q/2\cr0\cr0\cr -Q/2\cr},\
p'=\pmatrix{Q/2\cr0\cr0\cr Q/2\cr};
$$
on a remplac\'e l'\'energie de chaque pion
$$E_{\pi}={Q\over2}\left(\sqrt{1+{4m_{\pi}^2\over Q^2}}\right)$$
par sa valeur approch\'ee $Q/2$.

Pour d\'ecrire le pion dans son \'etat de valence, on introduit l'amplitude
de Bethe-Salpeter~(BS)~[Sal51]
$$
\langle0|T\left(q_{u\alpha i}(y)P_{ij}(y,0)\bar{q}_{d\beta
j}(0)\right)|\pi^+(p)\rangle,
\num
 $$
o\`u $u$ et $\bar{d}$ sont les saveurs des quarks de valence du $\pi^+$,
$\alpha$ et $\beta$ sont les indices de Dirac et $i$, $j$ les indices de
couleur. L'op\'erateur $P_{ij}(y,0)$ est n\'ecessaire pour rendre l'expression
compatible avec l'invariance dans les transformations de jauge locales: dans
une transformation $q(y)\rightarrow U(y)\,q(y)$, il se transforme selon
$P(y,0)\rightarrow U^{-1}(y)P(y,0)U(0)$ compensant ainsi les variations pour le
quark au point $y$ et l'antiquark \`a l'origine (nous reviendrons sur cette
question). La fonction d'onde de Bethe-Salpeter est la g\'en\'eralisation
relativiste de la fonction d'onde de Shr\"odinger pour d\'ecrire un \'etat
li\'e d'une paire quark antiquark~[Lurie~p.424]. On peut l'interpr\^eter comme
l'amplitude de probabilit\'e de trouver dans un pion $\pi^+$ un quark $u$ au
point $y$ et un antiquark $\bar{d}$ \`a l'origine.

En fait les r\`egles de Feynman sont habituellement donn\'ees en terme
d'impulsion, aussi d\'efinit-on la transform\'ee de Fourier de
l'amplitude~BS, soit
$$
\int d^4ye^{ik.y}<>=X_{\alpha\beta}(k,p-k)
\num
$$
o\`u $k$ est l'impulsion du quark et, par conservation de l'impusion,
$p-k$ celle de l'antiquark.

Pour discuter des propri\'et\'es de cette amplitude, il est int\'eressant
d'introduire les coordonn\'ees dites ``c\^one de lumi\`ere'' des impulsions
$$
\left\{\matrix{
k^+={1\over \sqrt{2}}(k^0-k^3)\cr
k^-={1\over \sqrt{2}}(k^0+k^3)\cr}\right.
\num
$$
Dans ce qui suit, on pr\'esentera les coordonn\'ees c\^one de lumi\`ere d'un
quadrivecteur sous la forme d'une liste \'ecrite entre crochets (pour la
distinguer de la liste des composantes habituelles entre parenth\`eses), selon
$k=[k^+,k^-,k^1,k^2]=[k^+,k^-,{\bf k}_{\bot}]$.
Le produit scalaire de deux quadrivecteurs A et B dans ces coordonn\'ees
s'\'ecrit
$$
A.B=A^+B^- +A^-B^+ -{\bf A}_{\bot}.{\bf B}_{\bot}.
\num
$$

On peut revenir aux quadrivecteurs du probl\`eme pour lesquels on trouve
$$
p=[Q/\sqrt{2},0,0,0],\ p'=[0,Q/\sqrt{2},0,0],
$$
puis param\'etrer les impulsions internes $k=[xQ/\sqrt{2},k^-,{\bf k}_{\bot}]$,
o\`u l'on a introduit la fraction c\^one de lumi\`ere $x$ que porte le quark
\`a l'int\'erieur du pion. On trouve facilement l'expression de l'impulsion
de l'antiquark et, en particulier, la fraction $1-x=\bar{x}$ qu'il porte. On
proc\`ede de m\^eme pour le pion final,
$k'=[k'^+,x'Q/\sqrt{2},{\bf k'}_{\bot}]$, \etc

En terme de ces variables, on peut alors d\'ecrire simplement les r\'egions
$k^{\mu}$ de l'impulsion interne favoris\'ee par l'amplitude $X(k,p-k)$:
$$
{\bf k}_{\bot}^2\lapprox M^2,\hskip 1cm |k^-|\lapprox M^2/Q.
$$

\section{Calcul du terme dur \`a l'approximation de Born}

L'\'el\'ement de matrice Eq.~(10) s'exprime comme la convolution,
repr\'esent\'ee sur la Figure~5,
$$
\int {d^4k\over (2\pi)^4}{d^4k'\over (2\pi)^4}\,
X(k)\,T_H^{\mu}(k,k')\,X^{\dag}(k').
\num
$$
\newcount\convolution\convolution\the\equano
A l'ordre le plus bas dans la constante de couplage $g$ de QCD, on trouve 4
diagrammes de Feynman. L'un d'eux est dessin\'e Figure~6 et les 3 autres s'en
d\'eduisent en attachant successivement le photon aux points 2, 3 et 4.

\midinsert
\vskip 3cm
{\noindent\sl Figure 6: Graphe en arbre pour le facteur de forme; les 3 autres
s'en d\'eduisent en attachant le photon aux points 2, 3 et 4. Les propagateurs
des lignes qui joignent les amplitudes de Bethe-Salpeter aux 3 vertex sont
absorb\'es, par d\'efinition, dans ces amplitudes.}
\endinsert

Evaluons, tout d'abord, le carr\'e de l'impulsion du gluon qui forme,
en jauge de Feynman, le d\'enominateur du propagateur du gluon. On a
$$\matrix{
(p'-k'-p+k)^2=&-\bar{x}\bar{x}'Q^2
&-\sqrt{2}Q(k^-\bar{x}'+k'^-\bar{x})&-2k^-k'^+
&-({\bf k}_{\bot}-{\bf k}'_{\bot})^2\cr
\noalign{\medskip}
&O(Q^2)&O(M^2)&O({\displaystyle M^4\over\displaystyle Q^2})&O(M^2)\cr
}
\num
$$
o\`u l'on a indiqu\'e les ordres de grandeurs typiques dans les r\'egions
d'impulsions favoris\'ees par les amplitudes $X(k)$ et $X(k')$. Comme
on se limite au calcul de l'amplitude dominante en $Q$, on oublie les termes
d'ordre $M^2$. Soit, en particulier,
$$
(p'-k'-p+k)^2\approx-\bar{x}\bar{x}'{\displaystyle Q^2\over\displaystyle 2}.
$$
L'analyse se r\'ep\`ete pour les autres quantit\'es qui interviennent
dans l'amplitude dure $T_H^{\mu}$, ce qui conduit \`a
$$
T_H^{\mu}(k,k')\approx
T_H^{\mu}\left(x{Q\over\sqrt{2}},x'{Q\over\sqrt{2}}\right).
\num
$$
On peut alors exprimer la convolution de l'\'equation~(13) sous la
forme
$$
\int dxdx'\,
\left({Q\over2\sqrt{2}\pi}\int {dk^-d{\bf k}_{\bot}\over (2\pi)^3}X(k)\right)
T_H^{\mu}(x,x')
\left({Q\over2\sqrt{2}\pi}
\int {dk'^+d{\bf k'}_{\bot}\over (2\pi)^3}X^{\dag}(k')\right),
\num
$$
et la quantit\'e n\'ecessaire \`a la description du pion dans cette r\'eaction
est, en fait, un objet beaucoup plus simple que l'amplitude $X$ en raison de
l'int\'egration sur trois des quatre composantes de l'impulsion interne.

Une premi\`ere simplification provient de l'int\'egration sur la variable
$k^-$ (pour le pion sortant c'est la variable $k'^+$).
En terme de la variable $y^+$ (conjugu\'ee de Fourier de $k^-$),
cela signifie que l'on se limite \`a l'amplitude de Bethe-Salpeter
en $y^+=0$ que l'on appelle fonction d'onde sur le c\^one de lumi\`ere et
que l'on note habituellement $\psi(x,{\bf k}_{\bot})$~[Bro89]. Une des
propri\'et\'es de cette fonction d'onde est d'avoir un support born\'e
dans la fraction $x$, soit $0\le x\le 1$.
La limitation aux fractions c\^one de lumi\`ere $x$ comprises entre 0 et 1
peut \^etre retrouv\'ee en exprimant $X$ sous la forme
$$
X(k,p-k)={f(k)\over [k^2-m^2+i\varepsilon]\ [(p-k)^2-m^2+i\varepsilon]},
$$
et en \'evaluant l'int\'egrale sur $k^-$ de $-\infty$ \`a $+\infty$ par la
m\'ethode de Cauchy. On n'obtient alors une contribution que lorsque les
deux p\^oles sont situ\'es de part et d'autre de l'axe r\'e\'el. Ces
p\^oles sont situ\'es en
$$\left\{\matrix{
k_1^-&=&{\sqrt{2}({\bf k}_{\bot}^2+m^2)\over xQ}
&-i\varepsilon\,{\rm sgn}(x)\hfill\cr
k_2^-&=&-{\sqrt{2}({\bf k}_{\bot}^2+m^2)\over \bar{x}Q}
&+i\varepsilon\,{\rm sgn}(1-x)\cr
}\right.$$
de telle sorte que l'int\'egrale produit un facteur\footnote{$^{\dagger}$}{
$\theta$ est la fonction d\'efinie par:
$$
\theta(x)=0, \hbox{ si }x<0,\hskip 1cm\theta(x)=1, \hbox{ si }x>0.$$}
$$
\theta(x)\theta(1-x);
$$
l'int\'egrale sur $x$ est alors limit\'ee \`a l'intervalle $[\,0,\,1\,]$.

En second lieu, on peut \'etudier la structure de Dirac de
l'amplitude $X(k)$ int\'egr\'ee sur $k^-$ et ${\bf k}_{\bot}$ (en terme
de coordonn\'ees, on a $y^+=0$ et ${\bf y}_{\bot}={\bf 0}$).
C'est une matrice $4\times 4$ qui
d\'epend du quadrivecteur $p$ et de la fraction $x$. On d\'efinit donc
$$
M_{\alpha\beta}(x,p)=
{Q\over2\sqrt{2}\pi}\int {dk^-d{\bf k}_{\bot}\over (2\pi)^3}X(k).
\num
$$
On trouve de plus que, compte-tenu
de la parit\'e $-1$ du pion, cette matrice a la propri\'et\'e
$$
M_{\alpha'\beta'}(x,\tilde{p})=
-\gamma^0_{\alpha'\alpha}M_{\alpha\beta}(x,p)\gamma^0_{\beta\beta'},
$$
o\`u $\tilde{p}^{\mu}=(p^0,-{\bf p})$. La combinaison la plus g\'en\'erale qui
satisfait cette identit\'e est
$$
M_{\alpha\beta}(x,p)=
{1\over4}\left[\gamma^5(p\slash\,\varphi(x)+\varphi'(x))\right]_{\alpha\beta}.
\num
$$
L'amplitude $\varphi'$ ne produit que des termes d'ordre $M^2/Q^2$ que
l'on n\'eglige devant l'amplitude due \`a $\varphi$\footnote{$^{\ddagger}$}{la
structure de Dirac ${1\over4}\gamma^5\,p\slash$
correspond \`a la combinaison des spineurs d'h\'elicit\'e ($\uparrow$ et
$\downarrow$ d\'esignent respectivement des \'etats d'h\'elicit\'e $+$ et
$-$)
$$
{1\over4}\gamma^5\,p\slash|_{\alpha\beta}=
{1\over2\sqrt{2x\bar{x}}}{1\over\sqrt{2}}\left(
u_{\alpha}(x{\bf p},\uparrow)\,\bar{v}(\bar{x}{\bf p},\downarrow)
-u_{\alpha}(x{\bf p},\downarrow)\,\bar{v}(\bar{x}{\bf p},\uparrow)\right).
$$

On retrouve la fonction d'onde de spin du pion du mod\`ele de quarks
$$
{1\over\sqrt{2}}\left(|\uparrow\downarrow\rangle
-|\downarrow\uparrow\rangle\right).
$$}.

La fonction $\varphi(x)$ est appel\'ee {\it amplitude de distribution}, elle
``mesure'' comment l'impulsion du pion est distribu\'ee entre son quark et son
antiquark de valence lorsque leur s\'eparation dans le plan transverse est
nulle.
C'est l'amplitude non perturbative qui connecte la physique de grande distance
de l'interaction forte \`a celle de petite distance.

Les approximations ci-dessus suivent la ligne de conduite que l'on s'est
fix\'ee : calculer l'amplitude dominante $A^{\rm LT}$ du
processus exclusif envisag\'e. A cette fin, on a isol\'e, d'une part, la partie
dominante de l'amplitude dure, et, d'autre part, le terme de la fonction
d'onde qui contribue \`a l'ordre dominant.

Il nous reste \`a pr\'eciser le traitement de la couleur. Un moyen de
simplifier
cette \'etude consiste \`a choisir, pour un pion volant selon la direction
$+$, les
jauges axiales d'axe le long de la direction $-$ (on fixe $A^+=0$). Pour
ces jauges,
on a $P_{ij}(y,0)=\delta_{ij}$. On peut en effet v\'erifier que les
transformations de
jauge compatibles avec $A^{a+}=0$ sont telles que $\partial^+\theta^a=0$, de
telle sorte que $\theta^a(y)$ est ind\'ependant de $y^-$. Ainsi
$$
\delta^{ij}u'^i(y^-)\bar{d}'^j(0)=
\delta^{ij}e^{igT^a\theta^a(y^-)}|^{ik}e^{-igT^a\theta^a(0)}|^{lj}u^k(y^-)\b
ar{d}^l(0)
=\delta^{ij} u^i(y^-)\bar{d}^j(0).
$$
Il s'ensuit que la composante de couleur pour la paire quark-antiquark est
simplement
$$
{1\over 3}\delta_{ij}.
$$
Ce fait explique, en partie, l'int\'er\^et des jauges c\^one de lumi\`ere
dans l'\'etude
des processus durs. Pour un autre choix de jauge, la forme explicite de
$P_{ij}(y,0)$
est n\'ecessaire, mais nous n'en parlons pas ici. Notons simplement que
$P_{ij}(y,0)$
peut \^etre analys\'e de mani\`ere perturbative et l'invariance de jauge
assur\'ee
ordre par ordre du d\'eveloppement perturbatif. A l'ordre $0$, qui va tout
d'abord
nous int\'eresser, on a
$$
P_{ij}(y,0)=\delta_{ij}+O(g).
$$

\smallskip

Nous sommes enfin arm\'es pour calculer le graphe de la Figure~6 avec la
r\`egle de Feynman pour le pion
$$
{1\over 3}\delta_{ij}{1\over4}\gamma^5\,p\slash|_{\alpha\beta}\,\varphi(x),
$$
et l'int\'egrale de boucle $\int_0^1dx$. L'amplitude du processus
(\the\convolution) peut donc \^etre r\'e\'ecrite
$$
\int_0^1dx\int_0^1dx'\varphi(x)\langle T_H^{\mu}(x,x')\rangle\varphi^*(x')
\num
$$
o\`u le processus dur est \'evalu\'e sur les composantes de couleur et de spin
\'ecrites ci-dessus. L'alg\`ebre de couleur conduit \`a la trace
$$
{1\over3}\delta_{ij}T^a_{jk}{1\over 3}{\delta_{kl}\over3}T^a_{li}
={C_F\over3}={4\over 9},
$$
et l'amplitude est, en n\'egligeant les masses des quarks,
$$\eqalign{
&\int_0^1dx\int_0^1dx'
(-){C_F\over3}Tr\left\{e_u\gamma^{\mu}{1\over4}\gamma^5p\slash\,
g\gamma^{\alpha}{1\over4}\gamma^5p\slash'\,g\gamma^{\beta}
{p'-\bar{x}p\over -\bar{x}Q^2}\right\}
{-\eta_{\alpha\beta}\over -\bar{x}\bar{x}'Q^2}\varphi(x)\varphi^*(x')\cr
&=
e_up^{\mu}{C_Fg^2\over6Q^2}\left|\int_0^1dx{\varphi(x)\over\bar{x}}\right|^2.
}
\num
$$
Le graphe o\`u le photon est attach\'e au point 2 conduit \`a la m\^eme
expression en changeant $p^{\mu}$ en $p'^{\mu}$. Enfin les deux autres graphes
sont identiques aux deux premiers aux \'echanges $e_u\leftrightarrow-e_d$ et
$\bar{x}\leftrightarrow x$ au d\'enominateur de l'int\'egrand pr\`es.
L'invariance par conjugaison de charge et la sym\'etrie d'isospin entrainant la
relation $\varphi(x)=\varphi(\bar{x})$, on peut mettre en facteur la
d\'ependance $(e_u-e_d)(p+p')^{\mu}$ attendue Eq.~(\the\parametrisation) et
isoler l'expression du facteur de forme
$$
F_{\pi}(Q^2)=
{C_Fg^2\over6Q^2}\left|\int_0^1dx{\varphi(x)\over\bar{x}}\right|^2.
\num
$$
\newcount\valeur\valeur\the\equano

Remarquons que l'on retrouve bien la loi d'\'echelle en $Q^{-2}$ pr\'edite
au moyen des r\`egles de comptage.
\smallskip

Quoiqu'on ait affirm\'e ne pas conna\^itre beaucoup de chose sur la fonction
d'onde de valence du pion, il existe quand m\^eme une contrainte fix\'ee par
la dur\'ee de vie du pion. Le processus est d\'ecrit sur la Figure~7.

\midinsert
\vskip 3cm
\centerline{\sl Figure 7: D\'esint\'egration du pion.}
\endinsert

On peut, comme pour le facteur de forme, isoler la transition faible au
niveau des quarks, sous la forme de l'\'el\'ement de matrice du courant
\'electrofaible~[Donoghue]. On a
$$
\langle0|\bar{q}_d(0)\gamma^{\mu}(1-\gamma^5)q_u(0)|\pi^+(p)\rangle
=f_{\pi}p^{\mu}, \num
$$
o\`u la constante de d\'ecroissance, $f_{\pi}$, vaut, dans cette
param\'etrisation, 133MeV.

L'amplitude (BS) \`a l'origine peut \^etre \'ecrite sous la forme
$$
\langle0|T\left(q_{u\alpha i}(0)\bar{q}_{d\beta j}(0)\right)|\pi^+(p)\rangle
=\int_0^1dx{Q\over2\sqrt{2}\pi}\int {dk^-d{\bf k}_{\bot}\over (2\pi)^3}X(k),
$$
que l'on peut multiplier par le tenseur
$[\gamma^{\mu}(1-\gamma^5)]_{\beta\alpha}\delta_{ji}$
$$
-\langle0|\bar{q}_{d i}(0)\gamma^{\mu}(1-\gamma^5)
q_{u i}(0)|\pi^+(p)\rangle
=Tr\left({\gamma^5p\slash\over4}\gamma^{\mu}(1-\gamma^5)\right)
{\delta_{ij}\over3}\delta_{ji}\int_0^1dx\,\varphi(x),
$$
o\`u l'on remarque que la composante $\varphi'$ ne survit pas \`a la
projection. Il vient, en utilisant l'Eq.~(\the\equano),
$$
p^{\mu}\int_0^1dx\,\varphi(x)=f_{\pi}p^{\mu},
\num
$$
ce qui fixe la normalisation de la distribution.

\section{Prise en compte des corrections radiatives}

Comme on l'a vu dans le cas du rapport $R$ (cours d'Eric Pilon), il est
important,
lorsque l'on calcule une quantit\'e au moyen de QCD perturbative, de
s'assurer que les corrections radiatives ne viennent pas bouleverser le
r\'esultat qui a \'et\'e obtenu \`a l'ordre le plus bas.

Le r\'egime ultraviolet ne pose pas de probl\`eme puisque la th\'eorie est
renormalisable. En fait les soustractions \`a effectuer pour n'importe quel
processus sont automatiquement prises en compte si on a trait\'e correctement
 les propagateurs de quarks et de gluons, d'une part, et le couplage,
d'autre part.

Le r\'egime infrarouge doit par contre \^etre envisag\'e avec attention. Pour
le processus examin\'e ici on trouve pour un diagramme \`a $n$ boucles des
corrections d'ordre
$$
{\alpha_S(Q^2)\over Q^2}\left[\alpha_S(Q^2)\ln {Q^2\over M^2}\right]^n,
$$
qui compte-tenu de $\alpha_S(Q^2)\propto(\ln Q^2/\Lambda^2)^{-1}$
 est de l'ordre du terme en arbre que l'on a calcul\'e !

On peut resommer ces grands logarithmes de telle mani\`ere \`a restaurer
la pr\'edictibilit\'e du formalisme. C'est ce que l'on appelle la
{\it factorisation} car le processus peut \^etre d\'ecrit par la convolution
illustr\'ee Figure~5 et \`a laquelle on \'etait arriv\'e Eq.(\the\convolution):
$$
F_{\pi}=\varphi*T*\varphi^*
$$
o\`u:
{\parindent 7mm
\item{--} $T$ est une amplitude dure que l'on peut l\'egitimement \'evaluer
au moyen de QCD perturbative; autrement dit, les corrections d'ordre
sup\'erieur \`a $T$ sont en $\alpha_S^n(Q)$, effectivement faibles \`a
suffisamment grand transfert;
\item{--} dans $\varphi$ sont absorb\'es tous les grands logarithmes; la
distribution $\varphi$, qui repr\'esente la fonction d'onde, d\'epend
d\'esormais de l'\'echelle $Q$ \`a laquelle le photon
virtuel sonde le pion. C'est une quantit\'e essentiellement non perturbative
puisqu'elle exprime la disposition que prennent les quarks de valence
confin\'es lorsqu'ils se pr\'esentent \`a petite distance pour interagir
dans un processus exclusif.
\par}

\smallskip
Examinons comment on resomme les logarithmes dominants (LL) pour construire
la distribution $\varphi_{\rm LL}$~[Field]. Pour se faire, il est
int\'eressant de choisir une jauge diff\'erente dela jauge de Feynman.
On se place dans la jauge axiale, d'axe $n^{\mu}$, en fixant la condition
sur le champ des gluons $A^{\mu}$: $n.A=0$. Avec ce choix, les corrections
dominantes prennent la forme illustr\'ee sur la Figure~8.

\midinsert
\vskip 3cm
\centerline{\sl Figure 8: Corrections dominantes en jauge axiale.}
\endinsert

On peut montrer que la sommation des graphes s'arrange de la fa\c con
suivante
$$\eqalign{
\varphi_{\rm LL}(x,Q)=&\,\varphi_0(x)
+\kappa\int_0^1du\,V_{q\bar{q}\rightarrow q\bar{q}}(u,x)\,\varphi_0(u)\cr
&+{\kappa^2\over2!}\int_0^1duV_{q\bar{q}\rightarrow q\bar{q}}(u,x)
\int_0^1du'V_{q\bar{q}\rightarrow q\bar{q}}(u',u)\,\varphi_0(u')+\ldots\cr
}
\num
$$
$\kappa$ contient les grands logarithmes colin\'eaires sous la forme
$$
\kappa={1\over\beta_1}\ln{\alpha_S(\mu^2)\over\alpha_S(Q^2)},\hbox{ o\`u}
\ \ \ \beta_1={1\over 4}(11-{2\over3}n_f).
$$
$V_{q\bar{q}\rightarrow q\bar{q}}$ est un noyau caract\'eristique
de la distribution de valence du pion
$$
V_{q\bar{q}\rightarrow q\bar{q}}(u,x)={2\over3}\left\{
{\bar{x}\over\bar{u}}\left(1+{1\over u-x}\right)_+\theta(u-x)+
{x\over u}\left(1+{1\over x-u}\right)_+\theta(x-u)\right\},
\num
$$
o\`u $()_+$ est une distribution qui indique que les divergences
infrarouges (ici \`a la limite $u\rightarrow x$) se compensent
entre les graphes b et c de la Figure~8, ceci est une cons\'equence de la
neutralit\'e de couleur du hadron.

L'\'equation pour $\varphi$ (on omet l'indice LL dor\'enavant) peut
\^etre r\'e\'ecrite sous la forme int\'egro-diff\'erentielle
$$
\left({\partial\varphi\over\partial\kappa}\right)_x=
\int_0^1du\,V(u,x)\,\varphi(u,Q),\num
$$
qui a pour solution g\'en\'erale
$$
\varphi(x,Q)=x(1-x)\sum_n\phi_n(Q)C_n^{(3/2)}(2x-1);
\num
$$
les polyn\^omes de Gegenbauer $C_n^{(m)}$ (g\'en\'eralisation des
polyn\^omes de Legendre) sont en effet tels que
$$
\int_0^1du\,u(1-u)\,V(u,x)\,C_n^{(3/2)}(2u-1)=A_nx(1-x)C_n^{(3/2)}(2x-1),
$$
avec $A_n$ des coefficients particuliers d\'ependants de $n$. Injectant la
solution dans l'\'equation, on trouve
$$
\phi_n(Q)=\phi_n(\mu)e^{A_n\kappa}=\phi_n(\mu)\left(
{\alpha_S(\mu^2)\over\alpha_S(Q^2)}\right)^{A_n/\beta_1},
\num
$$
et les exposants du d\'eveloppement commencent selon
$$
{A_0\over\beta_1}=0,\ {A_2\over\beta_1}=-0,62,\ \ldots
\num
$$
Les termes impairs du d\'eveloppement disparaissent car la distribution est
sy\-m\'e\-tri\-que sur l'intervalle $[\,0,\,1\,]$.

Formant l'int\'egrale
$$
\int_0^1dx\,\varphi(x,Q)=\phi_0(Q)\int_0^1dx\,x(1-x)={\phi_0\over6}=f_{\pi}
$$
on peut donner le d\'ebut du d\'eveloppement
$$
\varphi(x,Q)=6f_{\pi}x(1-x)+(\ln Q^2)^{-0.62}\,\Phi_2\,x(1-x)[5(2x-1)^2-1]
+\ldots
\num
$$
La distribution asymptotique ($Q\rightarrow\infty$) du pion est donc
$$
\varphi(x,Q\rightarrow\infty)\sim 6f_{\pi}x(1-x).
\num
$$
Cependant, ceci ne nous apprend rien du d\'eveloppement effectif aux
\'energies accessibles et les constantes $\Phi_2,\ldots\Phi_n$ nous sont
inconnues.

C'est la limite de ce qui peut \^etre effectu\'e au moyen de QCD perturbatif
en ce qui concerne la distribution $\varphi$, c'est \`a dire pour comprendre
la mani\`ere dont l'interaction forte ``fabrique'' un \'etat de valence. Pour
aller plus loin d'autres m\'ethodes doivent \^etre envisag\'ees et une
premi\`ere possibilit\'e est une \'etude exp\'erimentale d\'etaill\'ee des
r\'eactions exclusives \`a grand transfert. Il existe aussi des approches
th\'eoriques comme les calculs sur r\'eseaux ou les r\`egles de sommes de QCD
avec lesquels on \'evalue les {\it moments} de la distribution
$$
\int_0^1dx(2x-1)^2\varphi(x,\mu),\ \ldots
$$
Une telle \'etude a conduit Chernyak et Zhitnitsky~[Che84] \`a proposer la
distribution
$$
\varphi_{\rm cz}(x,Q^2)=6f_{\pi}x(1-x)\left\{1+[5(2x-1)^2-1]\left(
{\ln Q^2/\Lambda^2\over\ln Q_0^2/\Lambda^2}\right)^{-0.62}\right\},
\num
$$
avec $Q_0\approx500$MeV. On montre sur la Figure~9 la distribution propos\'ee
par Chernyak et Zhitnitski.
\midinsert
\vskip3cm
\centerline{\sl Figure 9: La distribution CZ et son \'evolution avec
l'\'echelle $\mu^2$.}
\endinsert

\section{Comparaison avec les donn\'ees exp\'erimentales}

Sur la Figure~10, on a report\'e les points exp\'erimentaux pour le facteur de
forme du pion ainsi que la valeur obtenue au moyen de
l'expression~(\the\valeur) avec les modifications discut\'ees dans la
pr\'esente section, soit
$$
Q^2F_{\pi}(Q^2)={8\pi\over9}\alpha_S(Q^2)
\left|\int_0^1dx{\varphi(x,Q^2)\over\bar{x}}\right|^2,\num
$$
en prenant pour distribution la forme asymptotique (ligne tiret\'ee), puis
celle propos\'ee par Chernyak et Zhitnitsky (ligne pleine).
Comme l'\'evolution logarithmique est trop faible pour \^etre observ\'ee dans
le domaine de $Q^2$ explor\'e on se contente de la valeur donn\'ee par
l'expression ci-dessus avec $\alpha_S(5{\rm GeV}^2)\approx0.3$.
La distribution asymptotique semble fournir une contribution trop faible
dans le domaine des transferts explor\'es alors que la distribution CZ
produit apparemment un r\'esultat satisfaisant.

\midinsert
\vskip3cm
{\sl Figure 10: Evolution avec $Q^2$ du facteur de forme du pion et
valeur obtenue avec l'\'equation (\the\equano) pour la distribution
asymptotique (ligne tiret\'ee) et la distribution de Chernyak et Zhitnitsky
(ligne pleine).}
\endinsert

\section{Degr\'es de libert\'e transverses}

Les cons\'equences ph\'enom\'enologiques du traitement pr\'ec\'edent ne
sont pas tout \`a fait satisfaisantes pour deux raisons. La premi\`ere est
la diff\'erence observ\'ee entre le facteur de forme de genre temps et de
genre espace. Le facteur de forme de genre temps du pion appara\^it
dans la r\'eaction d'annihilation
$$
e^+e^-\rightarrow \pi^+\pi^-,
$$
o\`u la param\'etrisation (\the\parametrisation) reste valable, pour une
annihilation en un seul photon virtuel, avec la seule modification
$$
\langle\pi^+(p')\pi^-(p)|J^{\mu}(0)|0\rangle=e_{\pi}(p'-p)^{\mu}F_{\pi}(q^2).
$$
Le facteur de forme est toujours une fonction de la variable $q^2$, mais
le quadrivecteur $q$ du photon virtuel est maintenant du genre temps.
On observe un rapport 2 en faveur de la r\'egion temps qui n'est pas explicable
dans la formulation donn\'ee ci-avant, celle-ci conduisant \`a un rapport
unit\'e.

La seconde difficult\'e provient du fait qu'une \'etude des corrections \`a une
boucle~[Dit81]
\`a l'expression de $T$ ci-dessus propose, pour \'eviter de larges corrections,
de fixer l'\'echelle du couplage $\alpha_S$ \`a la virtualit\'e $xx'Q^2$ du
gluon \'echang\'e plut\^ot qu'\`a celle du photon virtuel. Ce traitement
n'est correct que tant que le gluon est assez loin de sa couche de masse c'est
\`a dire tant que $x$ ou $x'$ n'approchent pas de 0. Or, num\'eriquement,
l'amplitude obtenue aux transferts interm\'ediaires provient en grande partie
de ces r\'egions. Si on ne veut pas rejeter enti\`erement l'image de petite
distance du processus \`a ces transferts, il faut r\'eexaminer les processus
\'el\'ementaires dans la r\'egion o\`u le gluon \'echang\'e devient mou.

Dans cette r\'egion les degr\'es de libert\'e d'impulsion transverse (ou de
distance transverse) deviennent importants et invalident l'approximation
colin\'eaire~[Li~92]. Avant d'examiner quantitativement l'effet de ces degr\'es
de libert\'e (dans la section D-2), expliquons qualitativement les
modifications
qui sont attendues.
L'interaction \'elastique d'un quark est supprim\'ee par un facteur de forme de
Sudakov qui quantifie la difficult\'e qu'a une charge acc\'el\'er\'ee \`a ne
pas
rayonner. De la m\^eme mani\`ere, l'interaction \'elastique d'un dip\^ole de
taille transverse $b$ est fortement supprim\'ee \`a moins que $b$ n'approche
$Q^{-1}$~[Bot89]. L'approximation qui consiste \`a n\'egliger les degr\'es de
libert\'e transverses nous place automatiquement dans la r\'egion
$b^2\lapprox (xx'|q^2|)^{-1}$, c'est \`a dire la r\'egion non-supprim\'ee
lorsque $xx'$ est d'ordre 1. Pour $xx'\rightarrow 0$, alors que cette
approximation devient ill\'egitime, on peut imaginer que la prise en compte
de la taille transverse permet, avec la suppression Sudakov associ\'ee,
d'\'eviter les contributions infrarouges dangereuses.

\chapter{C- G\'en\'eralisation aux autres processus}

Les r\'esultats obtenus pourle facteur de forme \'electromagn\'etique du pion
se g\'en\'eralisent aux autres processus exclusifs durs, avec une variante
importante pour les collisions hadron - hadron (voir Chapitre D). Ainsi, on
d\'efinit une amplitude de distribution pour le proton et on analyse le facteur
de forme magn\'etique $G_M$ de fa\c con tr\`es similaire. On peut ensuite
consid\'erer
des r\'eactions plus riches comme la diffusion Compton r\'eelle ou virtuelle.

\section{La fonction d'onde de valence du proton}

On peut comme pour le pion d\'efinir la fonction \`a trois points pour un
nucl\'eon
d'impulsion $p=[p^+,0,{\bf 0}]$ en jauge $A^+=0$. Avec ce dernier choix, on
trouve
facilement que la composante de couleur est $q^iq^jq^k\propto\varepsilon^{ijk}$
et restreindre notre \'etude \`a la projection sur le tenseur antisym\'etrique
$\varepsilon^{ijk}$:
$$
\int db_1^-db_2^-e^{ip^+(xb_1^-+yb_2^-)}\langle0|\varepsilon^{ijk}
q^i_{\alpha,f}(b_1^-)q^j_{\beta,g}(b_2^-)q^k_{\gamma,h}(0)|N(p,\lambda)\rangle.
\num
$$
L'antisym\'etrie de la fonction d'onde dans l'\'echange de deux quarks
implique que la projection ci-dessus est sym\'etrique dans l'\'echange des
indices
restants, c'est \`a dire
$$
\pmatrix{
x&\alpha&f\cr
y&\beta &g\cr
z&\gamma&h}
=\pmatrix{
y&\beta &g\cr
x&\alpha&f\cr
z&\gamma&h}.
$$

La sym\'etrie d'isospin nous permet de nous focaliser sur le proton dont le
terme de
saveur est une combinaison de $uud$, $udu$ et $duu$. On peut alors \'etudier la
composante $u_{\alpha}(x)u_{\beta}(y)d_{\gamma}(z)$ avant d'utiliser la
sym\'etrie ci-dessus pour obtenir les composantes $udu$ et $duu$.

La sym\'etrie restante entre les deux quarks $u$ est prise en compte  en
remarquant qu'avec la matrice de conjugaison de charges, $C$, qui est telle
que $C\gamma^{\mu}C=\,{^{T}\gamma^{\mu}}$ et $^TC=C^{-1}=-C$, on trouve
parmi les 16 matrices de Dirac: les 10 matrices sym\'etriques
$\gamma^{\mu}C,\,\sigma^{\mu\nu}C$ et les 6 antisym\'etriques
$C,\,\gamma_5C,\,\gamma_5\gamma^{\mu}C$.

Comme dans le cas du pion, on isole les composantes contenant le quadrivecteur
impulsion du hadron qui permettent de former les grands invariants
cin\'ematiques.
On trouve
$$\eqalignno{
\pmatrix{
x&\alpha\cr
y&\beta \cr
z&\gamma}
&=V(x,y,z)p\slash C\mid_{\alpha\beta} \gamma_5U\mid_{\gamma}
+A(x,y,z)p\slash\gamma_5 C\mid_{\alpha\beta} U\mid_{\gamma}\cr
&-iT(x,y,z)p_{\nu}\sigma^{\mu\nu}C\mid_{\alpha\beta}
\gamma_5\gamma_{\mu} U\mid_{\gamma},
\numalign
}$$
o\`u $U$ est solution de l'\'equation de Dirac $p\slash\,U=0$. Avec les
propri\'et\'es
de sym\'etrie des matrices $\gamma$, on d\'eduit que les amplitudes $V$ et $T$
(respectivement $A$) sont sym\'etriques (resp. antisym\'etrique)  dans
l'\'echange
de $x$ et $y$. $V$ et $A$ sont donc des combinaisons d'une fonction unique,
$\varphi(x,y,z)=V(x,y,z)+A(x,y,z)$. La forme de $U$ dans l'Eq.(\the\equano) est
fix\'ee par la condition sur l'h\'elicit\'e:
$\hat{p}\cdot\vec{J}\,\psi=\lambda\psi$,
o\`u l'op\'erateur d'h\'elicit\'e est $\hat{p}\cdot\vec{J}=
\hat{p}\cdot\vec{s}\otimes1_4\otimes1_4
+1_4\otimes\hat{p}\cdot\vec{s}\otimes1_4
+1_4\otimes1_4\otimes\hat{p}\cdot\vec{s}$
($\hat{p}$ est le vecteur unitaire le long de la direction de vol du proton,
$s_i={1\over4}\varepsilon_{ijk}\sigma^{jk}$ et $1_4$ est la matrice
d'identit\'e
$4\times4$ ). Avec la forme (\the\equano), on d\'eduit que
$\hat{p}\cdot\vec{s}\,U=\lambda U$, c'est \`a dire que $U$ est, en fait, le
spineur
du proton: $U(p,\lambda)$.

On peut maintenant construire
$$
\pmatrix{
x&\alpha&f\cr
y&\beta &g\cr
z&\gamma&h}=\delta_{fu}\delta_{gu}\delta_{hd}
\pmatrix{x&\alpha\cr y&\beta\cr z&\gamma}
+\delta_{fu}\delta_{gd}\delta_{hu}
\pmatrix{x&\alpha\cr z&\gamma\cr y&\beta}
+\delta_{fd}\delta_{gu}\delta_{hu}
\pmatrix{z&\gamma\cr y&\beta\cr x&\alpha}, \num
$$
et il reste \`a s'assurer que le proton est d'isospin ${1\over2}$ et non
${3\over2}$.
Cette condition est remplie si le terme sym\'etrique du tenseur de saveur
est nul,
c'est \`a dire
$$
\pmatrix{x&\alpha\cr y&\beta\cr z&\gamma}
+\pmatrix{x&\alpha\cr z&\gamma\cr y&\beta}
+\pmatrix{z&\gamma\cr y&\beta\cr x&\alpha}=0.
$$
On d\'eduit alors une derni\`ere relation entre $V$, $A$ et $T$ qui est
$$
2T(x,y,z)=\varphi(x,z,y)+\varphi(y,z,x),
$$
de telle fa\c con que, comme pour le pion, le nucl\'eon a une amplitude de
distribution unique que nous d\'etaillons ci-apr\`es.

L'amplitude de distribution du proton peut s'\'ecrire selon un d\'eveloppement
semblable \`a ce qui a \'et\'e \'ecrit plus haut pour le pion, mais cette
fois sur
des polyn\^omes diff\'erents :
$$
\varphi(x_i,Q)=120\,x_1 x_2 x_3\;\delta(x_1+x_2+x_3-1) \ \ \times
$$
$$
\left[1 + {{21}\over{2}}
\left({\alpha_S(Q^2)\over\alpha_S(Q_0^2)}\right)^{\lambda_1} A_1
 P_1(x_i)+
{{7}\over{2}} \left({\alpha_S(Q^2)\over\alpha_S(Q_0^2)}\right)^{\lambda_2}
 A_2 P_2(x_i) + \dots \right] ,
\num
$$
\newcount\proton\proton\the\equano
o\`u la lente \'evolution en $Q^2$ vient enti\`erement de facteurs
$ \alpha_S(Q^2)^{\lambda_i}$, o\`u les  $ \lambda_i$  sont des nombres
croissants
$$
\lambda_1 = {{5}\over{9\beta_1}},\hskip 1cm\lambda_2 = {{6}\over {9\beta_1}} ,
$$
et les $P_i(x_j) $ sont les polyn\^omes d'Appell:
$$
P_1(x_i)=x_1-x_3,\hskip 1cm P_2(x_i)=1-3x_2, \dots
$$
Les constantes $A_i$ sont inconnues et  mesurent la projection de la
fonction d'onde sur les  polyn\^omes d'Appell
$$
A_i=\int _{0}^{1}dx_1 dx_2 dx_3 ~\delta(x_1+x_2+x_3-1)~\varphi(x_i)~P_i(x_i)
$$

\section{Le facteur de forme magn\'etique du proton}

On peut de la m\^eme mani\`ere que pour le pion d\'ecrire l'interaction
\'elastique d'un proton et d'un \'electron
$$
e^-p\rightarrow e^-p,
$$
qui est plus ais\'ement accessible d'un point de vue exp\'erimental, au
moyen de deux facteurs de forme $F_1$ et $F_2$ (on fait toujours
l'hypoth\`ese de l'\'echange d'un seul photon virtuel entre \'electron et
proton)
$$
\langle p',h'|J^{\mu}(0)|p,h\rangle=e\bar{u}(p',h')\left[
F_1(Q^2)\gamma^{\mu}+i{\kappa\over2M}F_2(Q^2)\sigma^{\mu\nu}(p'-p)_{\nu}
\right]u(p,h);\num
$$
$h$ et $h'$ sont respectivement les h\'elicit\'es du proton entrant et
du proton sortant, $u$ et $\bar{u}$ leur spineur et $M$ est la masse du
proton. L'\'ecriture de l'\'el\'ement de matrice sous cette forme,
en particulier, le fait qu'il n'y a que deux facteurs de forme
\'electromagn\'etiques fonctions du seul scalaire $Q^2$, se d\'emontre
aussi au moyen des lois de sym\'etries g\'en\'erales~[Halzen].
Dans la d\'ecomposition ci-dessus, $e$ est l'oppos\'ee de la charge de
l'\'electron et $\kappa=1.79$ est le moment magn\'etique anormal du proton.
Avec ces conventions, les 2 facteurs de forme ont pour valeur \`a transfert
nul
$$
F_1(0)=1,\hskip3cm F_2(0)=1.
$$

Les deux facteurs de forme $F_1$ et $F_2$ sont respectivement appel\'es
facteur de forme de Dirac et Pauli. A partir de l'identit\'e de Gordon
$$
i(p'-p)_{\nu}\bar{u}'\sigma^{\mu\nu}u=
2M\bar{u}'\gamma^{\mu} u-(p+p')^{\mu}\bar{u}'u,
$$
on peut \'ecrire l'\'el\'ement de matrice du courant sous la forme
$$
\langle p',h'|J^{\mu}(0)|p,h\rangle=e\bar{u}'\left[
(F_1(Q^2)+\kappa F_2(Q^2))\gamma^{\mu}-{\kappa\over2M}F_2(Q^2)(p+p')^{\mu}
\right]u,
$$
ce qui permet d'introduire les facteurs de forme de Sachs qui apparaissent
dans la section efficace du processus; ce sont les combinaisons lin\'eaires
$$\eqalign{
G_M&=F_1+\kappa F_2\cr
G_E&=F_1+{q^2\over 4M^2}\kappa F_2.
}$$

Dans le pr\'esent formalisme (voir la section A-1), seul le facteur de forme
magn\'etique (dominant \`a grand transfert) est accessible. On obtient les
r\'esultats repr\'esent\'es sur la Figure~11.

\midinsert
\vskip3cm
{\sl Figure 11: Evolution avec $Q^2$ du facteur de forme magn\'etique du proton
{}~{\rm [Sil93]}}
\endinsert

\section{La diffusion Compton}

A l'ordre le plus bas dans la constante de structure fine $\alpha \sim {{1}
\over {137}}$,
la diffusion  Compton virtuelle  (VCS) est
d\'ecrite par la somme coh\'erente des amplitudes dessin\'ees sur la
Figure~12,
 \`a savoir le  processus de Bethe Heitler (Figure~12b) o\`u le  photon
final est
 rayonn\'e par l'\'electron et le vrai processus VCS (Figure~12a).
\midinsert
\vskip4cm

\centerline{\sl Figure~12: Amplitudes de la diffusion Compton virtuelle}
\endinsert
Comme l'amplitude BH est calculable \`a partir des
facteurs de forme \'elastiques  $G_{Mp}(-t)$ and $G_{Ep}(-t)$,
 son interf\'erence avec l'amplitude VCS
est une source d'information int\'eressante, absente de la diffusion
Compton r\'eelle . L'amplitude VCS d\'epend de trois invariants ind\'ependants;
 on choisit habituellement $ Q^2,  s, t $ ou $ Q^2, s, \theta_{CM} $.
\midinsert
\vskip4cm

\centerline{\sl Figure~13: La diffusion Compton r\'eelle \`a grand angle  }
\endinsert

La Figure 13  repr\'esente l'ensemble (pauvre) des donn\'ees disponibles
sur la diffusion Compton r\'eelle sur le proton avec
 $-t>1GeV^2$ [Shupe]. On a mis en ordonn\'ee
 $s^6 d\sigma/dt$  en fonction de  $\cos\theta_{CM}$
pour illustrer l'approche aux lois d'\'echelle asymptotiques.
Si on fait une interpolation des donn\'ees selon une loi en $s^{-\alpha}$,
on obtient
 $\alpha= 7.0\pm0.4$:
soit un \'ecart de  $ 2.5\ \sigma$ \`a la  pr\'ediction $\alpha=6$.

Le calcul perturbatif de la diffusion Compton r\'eelle~[Far88, Kro91] ou
virtuelle [Far91] commence par l'\'evaluation des 336 diagrammes
topologiquement distincts obtenus en couplant deux photons aux trois quarks
avec \'echange de deux gluons. Il existe de plus
 42 diagrammes avec couplage \`a trois gluons mais on trouve que leur
facteur de couleur est nul.
Chaque quark entrant (resp. sortant) porte une  fraction $x$ (resp $y$) de
la composante $+$ (resp $-$) de l'impulsion
de son  proton parent, ainsi que des composantes le long des trois
autres directions. Tant que ces fractions
$x$ ou $y$ sont d'ordre 1, il est l\'egitime de n\'egliger ces trois
 autres composantes dans le processus dur et on obtient
$$
A=\varphi_{(uud)}\otimes T_H(\{x\},\{y\})\otimes \varphi'_{(uud)}(1+O(M^2/t)),
\num
$$

\section{Une strat\'egie d'analyse des donn\'ees}

La premi\`ere fa\c con d'analyser les donn\'ees est de comparer les points
exp\'eri\-men\-taux \`a un calcul fait avec des amplitudes de distribution
issues d'un mod\`ele th\'eorique. Kronfeld and Ni\v zi\'c [Kro91] ont ainsi
calcul\'e la diffusion Compton r\'eelle avec diverses amplitudes de
distribution (Figure~13). On observe un bon pouvoir discriminatoire de la
section efficace Compton vis \`a vis de la quantit\'e non perturbative
$ \varphi(x)$ que nous voulons conna\^\i tre.

Une fa\c con d\'ependant moins des a-priori th\'eoriques
pour extraire la fonction d'onde des donn\'ees est d'\'ecrire la section
efficace diff\'erentielle comme la somme de termes
$$
A_i T_H^{ij}(\theta) A_j
$$
\noindent
o\`u on a utilis\'e la d\'ecomposition de l'amplitude de distribution sur les
polyn\^omes d'Appell et o\`u les $T_H^{ij}$ sont les  int\'egrales sur les
variables $x$ et $y$ de l'amplitude dure \`a un angle de diffusion $\theta$
donn\'e  multipli\'ee par les deux polyn\^omes d'Appell $A_i(x)$ et $A_j(y)$.
Les $T_H^{ij}$ ont des expressions un peu longues mais peuvent
\^etre tra\^it\'ees num\'eriquement.

D\'eterminer alors l'amplitude de distribution du proton
\`a partir des donn\'ees, revient alors \`a extraire par une m\'ethode de
maximum de vraisemblance les param\`etres $A_i$ en amputant la s\'erie
(Eq.~(\the\proton)) \`a ses $n$ premiers termes et en v\'erifiant que
l'inclusion
du terme $n+1$ ne modifie pas sensiblement la conclusion. On peut ensuite
explorer d'autres r\'eactions, par exemple la diffusion Compton virtuelle,
qui doit \^etre comprise avec la m\^eme collection de $A_i$.

\section{Autres processus }

La photo- et l'\'electro-production de m\'esons \`a grand angle permettront
de sonder les amplitudes de distribution des m\'esons $\pi$ et $\rho$ de la
m\^eme fa\c con. La production d'\'etats finals $ K \Lambda $ permettra de
sonder la production de quarks \'etranges, ce qui s\'electionne quelques
diagrammes des sous-processus durs. L'analyse de ces possibilit\'es reste
\`a faire si on excepte quelques travaux dans le cadre simplificateur du
mod\`ele des diquarks [Kro92].

La d\'esintegration des particules lourdes a aussi \'et\'e \'etudi\'ee
dans ce formalisme. Nous n'avons pas le temps d'en parler.

\chapter{D- Formalisme pour les sous processus ind\'ependants}

On a vu plus haut que le m\'ecanisme de collisions ind\'ependantes, dit de
Landshoff, violait les r\`egles de comptage. Il s'agit maintenant d'analyser
dans le cadre de la QCD ce processus et de comprendre s'il permet lui-aussi
de factoriser une quantit\'e du genre fonction d'onde sensible \`a la
dynamique du confinement d'une partie dure calculable perturbativement.
Apr\`es que des solutions partielles aient \'et\'e propos\'ees, en particulier
par Mueller~[Mue81], le formalisme a \'et\'e d\'evelopp\'e par Botts et
Sterman~[Bot89]. Un des r\'esultats importants de ces \'etudes est d'avoir
montr\'e que le m\'ecanisme de collisions ind\'ependantes domine
asymptotiquement le m\'ecanisme de petite distance dans les collisions
hadron-hadron. Soulignons n\'eanmoins qu'il est {\sl sous-dominant dans
le cas des r\'eactions de photo- et d'\'electro-production}.

\smallskip
On peut d'une mani\`ere g\'en\'erale \'ecrire une amplitude d'h\'elicit\'e
du processus  de diffusion \'elastique $ \pi - \pi $ d\'ecrit Figure~4 comme
$$
A=\int
\left\{\prod_{i=1}^4 {d^4k_i\over(2\pi)^4} X_{\alpha_i\beta_i}(k_i)\right\}
\left.H(\{k\})H'(\{p-k\})\right|_{\{\alpha\beta\}},\num
$$
o\`u l'on sous-entend, pour simplifier les notations, les indices
d'h\'elicit\'e \'eventuels des m\'esons ainsi que les indices de couleurs des
quarks et o\`u on a not\'e $\{k\}$ pour $k_1,k_2,k_3,k_4$.
On traite le cas d'une seule saveur de quark.

Dans l'Eq.~(\the\equano), $X(k)$ est l'amplitude de Bethe-Salpeter
$$
\int d^4ye^{ik.y}
\langle0|T\left(q_{\alpha}(y)P(y,0)\bar{q}_{\beta}(0)\right)|M(p)\rangle,
$$
alors que $H$ et $H'$ sont les amplitudes dures des sous processus, soit une
somme de graphes de QCD perturbative. Par convention, le graphe $H$ est celui
o\`u entre le quark issu du m\'eson num\'ero 1.

La premi\`ere \'etape consiste \`a simplifier l'expression pr\'ec\'edente.
A cette fin, on examine les r\'egions cin\'ematiques qui dominent dans
l'int\'egrale pr\'ec\'edente, soit \`a cause du comportement des
diff\'erentes amplitudes, soit du fait de la conservation de l'impulsion
au niveau des diagrammes durs
$$
(2\pi)^4\delta(\textstyle\sum_i k_i)\,(2\pi)^4\delta(\textstyle\sum_i p_i-k_i)
=(2\pi)^4\delta(\textstyle\sum_i k_i)\,(2\pi)^4\delta(\textstyle\sum_i p_i),
$$
(la conservation globale est, en fait, extraite dans la suite pour produire
l'amplitude de Feynman sous sa forme habituelle).

\section{Factorisation cin\'ematique}

Pour simplifier au maximum la cin\'ematique au niveau des constituants, il
est int\'eressant d'attacher \`a chaque m\'eson $M_i$ une base ``c\^one de
lumi\`ere'', $(v_i,v_i',\xi_i,\eta)$. Dans le centre de masse, on choisit
la direction de vol de $M_1$ comme axe $\hat{3}$. En notant $\theta$
l'angle de diffusion de $M_3$ mesur\'e par rapport \`a $\hat{3}$, on choisit
l'axe $\hat{1}$ de telle mani\`ere que la direction de vol de $M_3$
soit $\cos\theta\,\hat{3}+\sin\theta\,\hat{1}$. On peut alors \'etablir la
liste des vecteurs de base annonc\'es
$$\normalbaselineskip 20pt
\matrix{
v_1&\hskip -2mm=\hskip -2mm&
v_2'&\hskip -2mm=\hskip -2mm&
{\displaystyle1\over\displaystyle\sqrt{2}}(\hat{0}+\hat{3}) \hfill
&v_1'&\hskip -2mm=\hskip -2mm&
v_2&\hskip -2mm=\hskip -2mm&
{\displaystyle1\over\displaystyle\sqrt{2}}(\hat{0}-\hat{3}) \hfill\cr
\xi_1&\hskip -2mm=\hskip -2mm&
\xi_2&\hskip -2mm=\hskip -2mm&\hat{1} \hfill
&\eta&\hskip -2mm=\hskip -2mm&\hat{2} \hfill\cr
v_3&\hskip -2mm=\hskip -2mm&
v_4'&\hskip -2mm=\hskip -2mm&
{\displaystyle1\over\displaystyle\sqrt{2}}
        (\hat{0}+\sin\theta\,\hat{1}+\cos\theta\,\hat{3})\ \hfill
&v_3'&\hskip -2mm=\hskip -2mm&v_4&
\hskip -2mm=\hskip -2mm&
{\displaystyle1\over\displaystyle\sqrt{2}}
        (\hat{0}-\sin\theta\,\hat{1}-\cos\theta\,\hat{3}) \hfill\cr
\xi_3&\hskip -2mm=\hskip -2mm&
\xi_4&\hskip -2mm=\hskip -2mm&\cos\theta\,\hat{1}-\sin\theta\,\hat{3};\hfill\cr
}$$
\`a l'aide de ces vecteurs et \`a la limite o\`u on n\'eglige la masse des
m\'esons par rapport \`a $Q=\sqrt{s/2}$, l'impulsion des m\'esons s'\'ecrit
simplement $p_i=Qv_i$.

Les approximations cin\'ematiques [Bot89] consistent alors au remplacement
$$
H(\{k\})\approx H(\{xQ\})
$$
o\`u $x_i$ est la fraction d'impulsion du quark ou de l'antiquark $i$
qui entre ou sort du diagramme $H$. Une approximation \'equivalente
pr\'evaut pour $H'$. On approxime aussi
$$
\delta^{(4)}(k_1+k_2-k_3-k_4)\approx{\sqrt{2}\over |\sin\theta|Q^3}
\prod_{i=2}^4\delta(x_1-x_i)\,\delta(l_1+l_2-l_3-l_4),
$$
avec $l_i$ l'impulsion port\'ee par le quark ou l'antiquark $i$ selon la
direction $\eta$.
L'\'equation pr\'ec\'edente indique que toutes les fractions d'impulsions
dans $H$ sont identiques, on note $x$, la fraction unique r\'esultante, et
$\bar{x}=1-x$, la fraction dans $H'$.

On peut alors r\'earranger les int\'egrales dans l'\'equation~(\the\equano),
en introduisant le param\'etre d'impact
$$
2\pi\delta(l_i)=\int_{-\infty}^{+\infty}db\,e^{-i(l_3+l_4-l_1-l_2)b},
$$
et la fonction d'onde ``hybride'' d'un m\'eson se propageant selon la
direction $+$,
$$
{\cal P}_{\alpha\beta}(x,b)=
Q\int\!{dl\over2\pi}e^{ilb}\int\!{dk^-dk^1\over(2\pi)^3}
X_{\alpha\beta}(xQ,k^-,k^1,l),
$$
pour obtenir
$$
A(s,t)={\sqrt{2}Q\over2\pi|\sin\theta|}\int_0^1dx\int_{-\infty}^{\infty}db
\Big[H(\{xp\})H'(\{\bar{x}p\})\Big]_{\{\alpha\beta\}}
\prod_{i=1}^4 {{\cal P}_{\alpha_i\beta_i}(x,b;p_i)\over Q}.\num
$$

Chaque processus dur a une loi d'\'echelle en $Q^{-2}$, de sorte que la
loi d'\'echelle na\"ive pour l'amplitude de la r\'eaction est en
$\overline{|b|}Q^{-3}$, o\`u $\overline{|b|}$ est une moyenne typique de
la distance entre le quark et l'antiquark dans l'\'etat de valence des
m\'esons. On retrouve la violation des  r\`egles de comptage mentionn\'ee
au chapitre~1.

Soulignons que pour une convolution du type ``petite distance'', on aurait
\'ecrit
$$
A'=\prod_{i=1}^4\varphi_i(x_i)*T_H(\{x\}),
$$
$T_H$ consistant, \`a l'ordre le plus bas, en l'\'echange de trois gluons
durs pour des $x_i$ diff\'erents les uns des autres. Dans cette convolution,
un des gluons devient mou lorsque tous les $x_i$ sont \'egaux et $T_H$
connait une divergence infrarouge du type $\int d^4k/k^4$~[Mue81].

\section{Factorisation dynamique et corrections Sudakov}

On a d\'ej\`a signal\'e l'importance des corrections radiatives pour les
processus qui impliquent des hadrons: pour \'evaluer une section efficace
au moyen de la th\'eorie perturbative, il faut s'assurer que les r\'egions
infrarouges sont sous contr\^ole dans les int\'egrations de boucles.

La prise en compte des corrections modifie l'expression de l'amplitude du
processus, Eq.~(\the\equano), pour produire la forme suivante
$$
A(s,t)={\sqrt{2}Q\over2\pi|\sin\theta|}\int_0^1dx\,H(\{xp\})\,H'(\{\bar{x}p\})
\int_{-1/\Lambda}^{+1/\Lambda}db\,U(x,b,Q)
\prod_{i=1}^4{{\cal P}_i^{(0)}(x,b)\over Q},
\num
$$
\newcount\amplitude\amplitude\the\equano
o\`u le facteur $U$ contient les corrections. Comme on va le voir, ces
corrections
sont importantes car elles suppriment fortement l'int\'egrand dans la
r\'egion o\`u
le param\`etre d'impact $b$ est sup\'erieur \`a l'\'echelle $1/Q$. C'est le
ph\'enom\`ene
de Sudakov d\'ej\`a mentionn\'e plus haut. Par suite, l'amplitude ainsi
resomm\'ee
est de nouveau domin\'ee par une dynamique de petite distance.

Limitons-nous  \`a consid\'erer les corrections dominantes.
Ces corrections proviennent, en jauge axiale, des corrections sur les fonctions
d'onde. A l'ordre dominant, l'\'equation satisfaite par $\cal P$ est~[Bot89]
$$
{\partial\over\partial\ln Q}{\cal P}(x,b,Q)=-{1\over2}\left(
\int_{1/b}^{xQ}d\ln\mu'\,\gamma_K+\int_{1/b}^{\bar{x}Q}d\ln\mu'\,\gamma_K
\right){\cal P}(x,b,Q),
$$
o\`u
$$
\gamma_K(\mu')={C_F\over\beta_1\ln\mu'},
$$
dont la solution est
$$
{\cal P}(x,b,Q)={\cal P}^{(0)}(x,b)\,\exp-S(x,b,Q)\num
$$
o\`u
$$
S(x,b,Q)=\left({c\over4}\ln xQ(u-1-\ln u)+x\leftrightarrow\bar{x}\right),
\hbox{ avec }u(xQ,b)=-{\ln b\over\ln xQ}.
$$
$c=2C_F/\beta_1=32/27$ pour trois saveurs de quark.

On observe dans la forme g\'en\'erale de $\cal P$ la forte suppression des
grandes
distances transverses $b\gg 1/Q$ (suppression Sudakov) et un r\'egime sans
correction
(${\cal P}\approx{\cal P}^{(0)}$) autour de $b\sim 1/Q$. On pr\'esente sur la
Figure~14
la forme de $-S$ en fonction de $x$ et $b$.

\midinsert
\vskip 3cm
{\sl Figure 14: Exposant de la suppression Sudakov pour la fonction
d'onde ($-S(x,b,Q)$) pour $Q=2$ et $6\,${\rm GeV} ($\Lambda=200\,${\rm MeV})
en fonction de la taille transverse $b$ du dip\^ole (en {\rm fm}) et de la
fraction d'impulsion $x$ d'un des deux constituants.}
\endinsert

On voit que pour les valeurs interm\'ediaires de l'\'energie la suppression
n'est
effective que dans la r\'egion des grandes distances transverses avec cependant
une d\'ecroissance tr\`es rapide vers $-\infty$; \`a haute \'energie, par
contre,
la correction elle-m\^eme force le processus \`a \^etre de petite distance.
Rappelons que c'\'etait par un tout autre m\'ecanisme, pr\'ecis\'ement le
d\'eroulement du sous processus dur, que l'on avait justifi\'e le caract\`ere
``petite distance'' des processus exclusifs (voir le premier chapitre).
\smallskip

Profitons-en pour revenir sur les probl\`emes concernant l'\'etude du
facteur de
forme du pion aux \'energies accessibles que l'on avait soulign\'es \`a la
section B-5.
On peut, avec la d\'ependance de la fonction d'onde dans la distance
transverse,
envisager le calcul du terme dur avec ces degr\'es de libert\'e
suppl\'ementaires~%
[Li~92]. On trouve la forme approximative suivante
$$
T(-xx'q^2,b)\approx {2\over 3}\pi\alpha_S(t)C_F\,K_0(\sqrt{-xx'q^2}\,b),
$$
\newcount\termedur\termedur\the\equano
o\`u $K_0$ est une fonction de Bessel modifi\'ee.

On peut alors expliciter l'int\'er\^et de la prise en compte des degr\'es
de libert\'e
de distance transverse au regard des difficult\'es signal\'ees plus haut.
D'une part,
avec la prise en compte des corrections radiatives group\'ees dans la fonction
d'onde, Eq.~(\the\equano), et l'analyse par le groupe de renormalisation,
l'\'echelle pertinente pour le couplage $\alpha_S$ qui intervient dans
l'expression
de $T$ ci-dessus est
$$
t=\max(1/b,\sqrt{xx'|q^2|}).
$$
Avec le r\'egime de suppression que l'on vient de d\'ecrire, on d\'eduit
que le facteur
de forme n'obtient de contribution importante, \`a grand transfert
($>5\,$GeV$^2$),
que dans la r\'egion o\`u $b$ est relativement petit de sorte que l'\'echelle
$t$ de l'approche perturbative reste suffisante sur tout le domaine de
contribution.

D'autre part, la forme de la fonction de Bessel $K_0$ est assez
diff\'erente dans les
divers domaines du plan complexe, et, en particulier, lorsque $q^2$ est
r\'eel, entre
$q^2>0$ et $q^2<0$. On peut donc envisager des diff\'erences lors du
passage de la
r\'egion espace \`a la r\'egion temps~[Gou95].
\smallskip

Revenons maintenant au processus de Landshoff. Le facteur $U$ dans l'expression
de l'amplitude, Eq.~(\the\amplitude),  est le produit des facteurs $e^{-S}$
provenant
des quatre fonctions d'onde, soit
$$
U(x,b,Q)=\exp-\left(c\ln xQ(u-1-\ln u)+x\leftrightarrow\bar{x}\right).\num
$$
\newcount\suppression\suppression\the\equano

Nous ne chercherons pas ici \`a expliciter le calcul des diagrammes durs
qui permet
de trouver la forme quantitative de l'amplitude, mais simplement \`a voir
comment
la pr\'esence du terme de suppression $U$ modifie la r\`egle de comptage
trouv\'ee
dans la section A-3.

A cette fin, on \'etudie le comportement de l'amplitude pour
$Q\rightarrow\infty$.
Aux grandes valeurs de $Q$, on peut \'evaluer analytiquement l'int\'egrale
en $b$
$$
\int_0^{\Lambda^{-1}}db\,U(b,x,Q),
$$
par la m\'ethode du col. Pour ceci on approxime l'exposant dans
l'expression de $U$
$$
c\ln xQ\left(-{\ln b\over\ln xQ}-1-\ln -{\ln b\over\ln xQ}\right)
+x\leftrightarrow\bar{x}
\approx 2c \ln\sqrt{x\bar{x}}Q\,(u-1-\ln u)
$$
o\`u $u=-\ln b/\ln\sqrt{x\bar{x}}Q$. On effectue alors le changement de
variable $b\rightarrow u$, soit
$$
\ln\sqrt{x\bar{x}}Q\int_0^{+\infty}du\,
\exp-\ln\sqrt{x\bar{x}}Q\left(2c(u-1-\ln u)+u\right)
$$
L'exposant est maximum pour $u_0={2c \over 2c+1}$ et on obtient
la valeur approximative
$$
\int_0^{\Lambda^{-1}}db\,U(b,x,Q)\approx
u_0\,\sqrt{\pi \ln Q \over c} (x\bar{x} Q^2)^{c \ln u_0}.
$$
L'int\'egration en $x$ ne modifie ce comportement que par des logarithmes.
On voit
donc que l'effet des corrections radiatives est de supprimer fortement la
contribution
au canal exclusif \'etudi\'e lorsque le param\'etre d'impact $b\gg 1/Q$. Ceci
se traduit par une contribution effective des seules r\'egions o\`u les
interactions ind\'ependantes sont suffisamment proches et par la modification
de la loi d'\'echelle $Q^{-3}\rightarrow Q^{-3.83}$, qui donne pour la
section efficace
diff\'erentielle $s^{-5}\rightarrow s^{-5.83}$. Dans le cas de la collision
\'elastique
proton-proton la modification est $s^{-8}\rightarrow s^{-9.7}$~[Bot89].

On trouve donc des lois d'\'echelle tr\`es proches de celles obtenues avec les
diagrammes des r\`egles de comptage ($s^{-10}$ dans le cas $p\,p$ \'elastique).
Ces deux types de processus sont d\`es lors \`a m\^eme d'\^etre en
comp\'etition
sur un intervalle d'\'energie important autour de l'\'energie \`a laquelle
ils ont la
m\^eme amplitude.  On en d\'eduit alors un sch\'ema naturel pour
expliquer~[Pir82] les oscillations de la section efficace diff\'erentielle
observ\'ees exp\'erimentalement comme dues \`a l'interf\'erence des amplitudes
de ces deux processus, le facteur de forme de Sudakov s'accompagnant d'une
phase "chromo-coulombienne" d\'ependant logarithmiquement de l'\'energie.

\chapter{E- La transparence de couleur}

La transparence de couleur, c'est l'histoire du passe-muraille appliqu\'ee \`a
 la physique nucl\'eaire. Tout le monde sait que l'interaction forte est forte
et donc qu'un proton ne peut pas traverser un noyau sans \^etre absorb\'e...Et
pourtant, si quelqu'un pouvait s'\'ecrier: {\it ch\'erie, j'ai r\'etr\'eci
le proton},
alors ce {\it mini-proton} pourrait librement se faufiler \`a travers la
mati\`ere
nucl\'eaire. Cette propri\'et\'e est en fait bien connue en
\'electrodynamique: la
 section efficace d'interaction d'un dip\^ole \'electrique est
proportionnelle au
carr\'e de sa taille, et ceci est d\^u au fait que l'interaction
\'electromagn\'etique
\'el\'ementaire est proportionnelle \`a la charge des objets. La
chromodynamique
quantique \'etant elle aussi une th\'eorie de jauge, dont la charge est la
couleur
selon SU(3), la section efficace d'un trip\^ole neutre de couleur comme un
triplet
blanc de trois quarks color\'es, est aussi proportionnelle au carr\'e de
la taille
caract\'eristique de cet objet.

Si maintenant une exp\'erience s\'electionne dans la fonction d'onde
du proton
les configurations les plus simples o\`u les distances caract\'eristiques
sont petites
et contr\^olables, le {\sl rapport de transparence} d\'efini comme la
section efficace de ce processus sur un noyau divis\'e par la somme des
sections efficaces sur des nucl\'eons
libres , doit s'approcher de 1. La proximit\'e de ce rapport \`a l'unit\'e
d\'epend de la taille r\'eduite  du mini-proton et de son \'evolution
pendant la travers\'ee du noyau consid\'er\'e.

Comme on l'a  vu plus haut, les r\'eactions exclusives dures s\'electionnent
les
\'etats de Fock minimaux dans la fonction d'onde du proton et seulement ceux
dont la distance entre quarks est de l'ordre de $ 1/Q $ , si $Q$ est
l'impulsion
 transf\'er\'ee. Tester l'id\'ee de la
transparence de couleur semble donc simple: il suffit de r\'ealiser une
exp\'erience exclusive sur un proton libre  puis sur des noyaux lourds et de
mesurer le rapport de
transparence comme une fonction de l'\'energie transf\'er\'ee.

Il n'est pas question de d\'evelopper le sujet de la transparence de couleur
qui \`a lui-seul m\'eriterait un cours. On pourra se reporter \`a une revue
r\'ecente~{\rm [Jai95]}. Nous nous contenterons ici de pr\'esenter les
quelques r\'esultats exp\'erimentaux disponibles actuellement.

Les premi\`eres donn\'ees exp\'erimentales ayant fait sortir la transparence de
couleur du domaine des sujets acad\'emiques sont venues de Brookhaven, dans la
r\'eaction pp \'elastique \`a $90^\circ$  (CM) dans le milieu
nucl\'eaire~{\rm [Car88]}.
Ces donn\'ees ont \'et\'e l'occasion d'un d\'ebat sans conclusion unanime. La
difficult\'e vient de la difficile compr\'ehension d\'etaill\'ee des
r\'eactions
exclusives hadroniques qui, on l'a vu plus haut sont riches en subtilit\'e
venant
de la sensibilit\'e infrarouge des processus de collisions ind\'ependantes.
Ceci
implique que certaines configurations du proton de taille transverse moins
petite contribuent aussi \`a la section efficace exclusive. Le ph\'enom\`ene de
transparence de couleur est donc remplac\'e par celui de {\it filtre
nucl\'eaire}:
la diffusion \'elastique \'elimine la composante \'epaisse de la fonction
d'onde
du nucl\'eon et donc sa contribution \`a la section efficace. Si on admet
que dans
l'interaction \'elastique des protons les deux processus possibles
interf\`erent
pour produire les oscillations observ\'ees dans la variation de la section
efficace diff\'erentielle avec l'\'energie (\`a angle fix\'e), on en d\'eduit
ais\'ement que le rapport de transparence doit osciller en opposition de
phase~{\rm [Ral88]}.
C'est ce que les donn\'ees exp\'erimentales sugg\`erent (voir Figure~15).
On aimerait \'evidemment que des donn\'ees \`a plus haute \'energie nous fasse
voir une deuxi\`eme oscillation pour conforter cette interpr\'etation. On
devrait bient\^ot les avoir.
\midinsert
\vskip 5cm

{\sl Figure~15: Oscillations du rapport de transparence
 dans la diffusion \'elastique $pp$  \`a $90^\circ$ }
\endinsert
Pour comprendre les donn\'ees, il est bon de d\'efinir une probabilit\'e de
survie reli\'ee de fa\c con standard \`a une section efficace effective
d'att\'enuation $\sigma_{\rm eff}(Q^2) $ et d'\'etudier la variation avec le
transfert  de cette section efficace d'att\'enuation~{\rm [Jai93]}. On obtient
de fait des valeurs de $\sigma_{\rm eff}(Q^2) $ qui d\'ecroissent avec $Q^2 $
et qui sont
bien inf\'erieures aux sections in\'elastiques habituelles. On trouve m\^eme
que la probabilit\'e de survie ob\'eit \`a une simple loi d'\'echelle en
$Q^2/A^{1/3}$~{\rm [Pir91]}.

 L'exp\'erience NE18 au SLAC~{\rm [SLAC]} a r\'ecemment mesur\'e le rapport de
transparence jusqu'\`a $Q^2=7 GeV^2$ sans observer de croissance significative.
Ces donn\'ees, repr\'esent\'ees sur la Figure 16 mettent en doute les visions
les plus optimistes de la dominance tr\`es pr\'ecoce des petites
configurations.
Elles mettent l'accent sur la n\'ecessit\'e d'un boost suffisant qui emp\`eche
les petites configurations de se rhabiller trop vite, ce qui ob\`ere leur
capacit\'e \`a traverser sans \^etre absorb\'ees par le noyau.

\midinsert
\vskip 5cm

\centerline{\sl Figure~16: Le rapport de transparence mesur\'e \`a SLAC }
\endinsert

L'\'electroproduction diffractive de m\'esons  vectoriels \`a
 Fermilab  a r\'ecemment montr\'e une croissance
remarquable du rapport de transparence pour des  $Q^2 \simeq 7
 GeV^2$ ~[Ada95].
Dans ce cas le boost est important puisque l'\'energie du  lepton est
$E \simeq 200 GeV$ mais le probl\`eme est d'\^etre s\^ur de distinguer
les \'ev\`enements diffractifs des r\'eactions in\'elastiques.
\midinsert
\vskip 5cm
\centerline{\sl Figure~17:  Le rapport de transparence mesur\'e \`a FNAL}
\centerline{\sl dans l'\'electroproduction  diffractive de $\rho$}
\endinsert

Des r\'eactions plus simples s'av\`erent indispensables \`a l'\'etude de la
transparence de couleur. L'absorption d'un photon virtuel par un proton d'un
noyau, sans \'emission de m\'esons et sans fission compl\`ete du noyau semble
la r\'eaction id\'eale de par sa simplicit\'e. Mais ce processus est
malheureusement tr\`es rare et son \'etude n'a pas pu \^etre men\'ee \`a bien
avec les acc\'el\'erateurs existants, malgr\'e l'essai courageux au SLAC
mentionn\'e plus haut. Il apparait en effet qu'une grande luminosit\'e, un
grand cycle utile et une tr\`es bonne r\'esolution en \'energie soient les
ingr\'edients indispensables d'une recherche pertinente~[ELFE].

Apr\`es qu'on ait v\'erifi\'e que les facteurs de forme
\'electromagn\'etiques  d\'ecroissent en $Q^2$  conform\'ement aux
pr\'edictions
 de QCD, les exp\'eriences
$eA \rightarrow e' (A-1)  ~ p$
mesureront les propri\'et\'es d'\'ecrantage de couleur de la th\'eorie. La
quantit\'e \`a mesurer est le rapport de transparence d\'efini par:
$$
T_r = {\sigma_{\rm Noyau}\over Z \sigma_{\hbox{\sevenrm Nucl\'eon}}}
\num
$$

On a aussi pour ce rapport une loi d'\'echelle qui relie la taille
du noyau et la virtualit\'e du photon. A grand $Q^2$, $T_r$
ne doit d\'ependre que de  $A^{1 \over 3}/Q^2$  ~{\rm [Pir91]}.
L'approche \`a ce comportement et la valeur de $T_r$ en fonction de cette
nouvelle variable determinent l'\'evolution  de la
configuration ponctuelle au hadron.

Etablir l'existence du ph\'enom\`ene n'est qu'une premi\`ere \'etape.
L'approche
\`a la transparence de couleur est en effet au moins aussi int\'eressante
puisqu'elle d\'epend crucialement de la dynamique du confinement
des quarks dans les protons. Prenons l'exemple d'un mini-proton qui vient
d'absorber un photon tr\` es virtuel. Pendant qu'il traverse la mati\`ere
nucl\'eaire \`a la mani\`ere d'un passe muraille, sa fonction d'onde
\'evolue vers son \'etat habituel , un \'etat qui n'autorise pas une telle
libert\'e de mouvement. Selon son \'energie, cette \'evolution se fera plus ou
moins vite et , comme le h\'eros de Marcel Aym\'e, il restera ou non
prisonnier.
L'\'etat actuel de la th\'eorie des interactions fortes dans son domaine non
perturbatif ne permet certes pas de pr\'edire pour chacun des processus
envisageables la d\'ependance en \'energie de leur rapport de transparence.
Mais il est clair que des donn\'ees exp\'erimentales pr\'ecises contraindraient
tout mod\`ele non perturbatif de l'\'evolution d'un mini-proton \`a un
proton, et donc de la dynamique du confinement.

\biblio{

\bibitem{Ada95}
M.R.Adams {\it et\ al}, Phys. Rev. Lett. {\bf 74} , 1525 (1995)

\bibitem{BNL850} {\it BNL--Experiment 850}; spokesmen A. Carroll and S.
Heppelmann

\bibitem{Bot89}
J. Botts et G. Sterman, Nucl. Phys. {\bf B325}, 62 (1989)

\bibitem{Bro73}
S.J. Brodsky et G.R. Farrar, Phys. Rev. Lett. {\bf 31}, 1153 (1973);
V.A. Matveev, R.M. Muradyan et A.V. Tavkhelidze, Lett. Nuovo
Cimento {\bf 7}, 719 (1973)

\bibitem{Bro89}
S.J. Brodsky et G.P. Lepage, {\sl Perturbative QCD} p.93, ed. par A.H. Mueller
(World Scientific, Singapour, 1989)

\bibitem{Car88}
 A.S. Carroll {\it et\ al}, Phys. Rev. Lett. {\bf 61}, 1698 (1988)

\bibitem{Che84}
V.L. Chernyak et A.R. Zhitnitsky, Phys. Rep. {\bf 112}, 173 (1984)

\bibitem{Dit81}
F.-M. Dittes et A.V. Radyushkin, Sov. J. Nucl. Phys. {\bf 34}, 293 (1981)

\bibitem{Donoghue}
J.F. Donoghue, E. Golowich et B.R. Holstein, {\sl Dynamics of the
Standard Model} (Cambridge University Press, Cambridge, 1992)

\bibitem{ELFE}  {\it The ELFE Project,} Conference Proceedings,
 Vol.44, Italian Physical Society, Bologna, Italy (1993) edited by J.
Arvieux and E.DeSanctis;
J. Arvieux and B. Pire, Progress in Particle and Nuclear Physics, {\bf 30},
 299 (1995)

\bibitem{Far79}
G.R. Farrar et D. Jackson, Phys. Rev. Lett. {\bf 43}, 246 (1979);
S.J. Brodsky et G.P. Lepage, Phys. Lett. {\bf 87B}, 359 (1979);
A.V. Efremov et A.V. Radyushkin, Phys. Lett. {\bf 94B}, 245 (1980);
V.L. Chernyak, V.G. Serbo et A. R. Zhitnitsky,
Yad. Fiz. {\bf 31}, 1069 (1980);
A. Duncan and A.H. Mueller, Phys. Rev. D {\bf 21}, 1636 (1980)

\bibitem{Far88}
G.R. Farrar and E. Maina, Phys  Lett {\bf B206} , 120 (1988);
Glennys R. Farrar and Huayi Zhang, Phys Rev Lett {\bf 65}, 1721 (1990);
Glennys R. Farrar and Huayi Zhang, Phys Rev {\bf D41}, 3348 (1990)

\bibitem{Far91}
G.R. Farrar, K. Huleihel et H. Zhang, Nucl. Phys. {\bf B349}, 655 (1991)

\bibitem{Field}
R.D. Field, {\sl Applications of Perturbative QCD}
(Addison-Wesley, Redwood, 1989)

\bibitem{Gou95} T. Gousset and B. Pire, Phys Rev D {\bf 51}, 15 (1995)

\bibitem{Halzen}
F. Halzen et A.D. Martin, {\sl Quarks and Leptons} p.175,
Wiley (1984, New York)

\bibitem{HERMES}
 {\it Hermes proposal}, Report DESY--PRC 90/01

\bibitem{Itzykson}
C. Itzykson and J.B. Zuber, {\sl ``Quantum Field Theory''}, 692
(McGraw-Hill, Singapore, 1985)

\bibitem{Jai93} P. Jain and J.P. Ralston , Phys. Rev. {\bf D48 }, 1104
 (1993) and in {\it Proceedings of the XXVIIth Rencontre de
Moriond on Elementary Particle Physics}, (Les Arcs, France 1993) edited by
J. Tran Thanh Van (Editions Frontieres, Gif-sur-Yvette, France 1993)

\bibitem{Jai95} for a long list of references, see  P. Jain,
 B. Pire and J.P. Ralston, to be published in Physics Reports; see also
 L. Frankfurt, G.A. Miller, and M. Strikman, Comments Nucl. Part. Phys.
{\bf 21}, 1 (1992)

\bibitem{Kro92}
 P. Kroll, M.Schurmann and W. Schweiger, {\it Proceedings of Quark Cluster
 Dynamics}, (Bad Honnef, Germany 1992), edited by K. Goeke (Springer-Verlag,
1992) p 179; M.Anselmino {\it et al.} Rev. Mod. Phys.66, 195 (1993)

\bibitem{Kro91}
A.S. Kronfeld and B. Ni\v zi\'c, Phys Rev {\bf D44}, 3445 (1991)

\bibitem{Lan74}
 P.V. Landshoff, Phys. Rev. D {\bf 10}, 1024 (1974)

\bibitem{Leader}
E. Leader et E. Predazzi, {\sl Gauge Theories and the New Physics}
(Cambridge University Press, Cambridge, 1982)

\bibitem{Li~92}
H.-N. Li et G. Sterman, Nucl. Phys. {\bf B381}, 129 (1992)

\bibitem{Lurie}
D. Lurie, {\sl Particles and Fields} (Wiley-Interscience, New York, 1968)

\bibitem{Mue81}
 A.H. Mueller, Phys. Rep. {\bf 73}, 237 (1981)

\bibitem{Pir82}
B. Pire and J.P. Ralston, Phys. Lett. {\bf B117}, 233 (1982)

\bibitem{Pir91}
B. Pire and J.P. Ralston, Phys. Lett. {\bf B256}, 523 (1991)

\bibitem{Ral88}
 J. P. Ralston and B. Pire, Phys. Rev. Lett. {\bf 61}, 1823 (1988)

\bibitem{Sal51}
E.E. Salpeter et H.A. Bethe, Physical Review {\bf 84}, 1232 (1951)

\bibitem{Shupe}
M.A. Shupe et al., Phys Rev {\bf D19}, 1929 (1979)

\bibitem{Sil93}
A.F. Sill {\it et al.}, Phys. Rev. D {\bf 48}, 29 (1993)

\bibitem{SLAC}
N. Makins {\it et al.}, Phys. Rev. Lett. {\bf 72}, 1986 (1994);

 T. G. O'Neill {\it et al.}, Phys. Lett., {\bf B351}, 87 (1995).
}

\bye